\documentclass[12pt,draftcls,onecolumn]{IEEEtran}
\usepackage{amsfonts}
\usepackage{amsthm}
\usepackage{amssymb}
\usepackage{epsfig}
\usepackage{graphicx}
\usepackage{times}
\usepackage[usenames,dvipsnames]{pstricks}
\usepackage{epsfig}
\usepackage{pst-grad} 
\usepackage{pst-plot} 
\usepackage{epstopdf}
\usepackage{color}
\usepackage{authblk}
\usepackage{geometry}
\usepackage{dsfont}
\usepackage{amsmath}
\def\Prob{{\rm Pr}\,}
\def\diag{{\rm diag}\,}

\def\Exp{{\mathbb{E}}\,}
\def\tr{{\mathrm tr}\,}

\def\diag{{\rm diag}\,}

\newcommand{\mb}[1]{\mathbf{#1}}
\newcommand{\mr}[1]{\mathrm{#1}}
\newcommand{\mc}[1]{\mathcal{#1}}

\def\br{{\bf r}}
\def\bu{{\bf u}}

 \def\bW{{\bf W}}

\def\bg{{\bf g}}
\def\bx{{\bf x}}
 \def\bY{{\bf Y}}
\def\vec{\mr{vec}}
\def\t{\mr{T}}

\newtheorem{theorem}{\textbf{Theorem}}
\newtheorem{proposition}{\textbf{Proposition}}

\newtheorem{lemma}{\textbf{Lemma}}

\newtheorem{remark}{\textbf{Remark}}

\newcommand{\kl}{^{[kl]}}
\newcommand{\kj}{^{[kj]}}

\newcommand{\ks}{^{[ks]}}

\newcommand{\nv}{^{[nv]}}

\newcommand{\sumvL}{\sum_{v=1}^L}
\newcommand{\sumjL}{\sum_{j=1}^L}

\newcommand{\sumsL}{\sum_{s=1}^L}

\title{Interference Reduction in Multi-Cell Massive MIMO Systems II: Downlink Analysis for a Finite Number of Antennas}
\date{}
\author[1]{Liangbin Li}
\author[2]{Alexei Ashikhmin}
\author[2]{Thomas Marzetta\thanks{
This work was done with the first author's summer internship with Bell Labs in 2012. Part of this work was presented at IEEE Allerton 2013.}
}
\affil[1]{University of California, Irvine, CA 92617.}
\affil[2]{Bell Laboratories Alcatel-Lucent, 600 Mountain Ave, Murray Hill, NJ 07974.}

\begin{document}
\maketitle
\begin{abstract}
Sharing global channel information at base stations (BSs) is commonly assumed for downlink multi-cell precoding. In the context of massive multi-input multi-output (MIMO) systems where each BS is equipped with a large number of antennas, sharing instant fading channel coefficients consumes a large amount of resource. To consider practically implementable methods, we study in this paper interference reduction based on precoding using the large-scale fading coefficients that depend on the path-loss model and are independent of a specific antenna. We focus on the downlink multi-cell precoding designs when each BS is equipped with a {\em{practically finite}} number of antennas. In this operation regime, pilot contamination is not the dominant source of interference, and mitigation of all types of interference is required. This paper uses an optimization approach to design precoding methods for equal qualities of service (QoS) to all users in the network, i.e.,maximizing the minimum signal-to-interference-plus-noise ratios (SINRs) among all users. The formulated optimization is proved to be quasi-convex, and can be solved optimally. We also propose low-complexity suboptimal algorithms through uplink and downlink duality. Simulation results show that the proposed precoding methods improve $5\%$ outage rate for more than $10^3$ times, compared to other known interference mitigation techniques.


\end{abstract}
\newpage
\section{Introduction}
Interference limits the capacity of cellular networks. As each base station (BS) increases transmit power, the received signal-to-interference-plus-noise ratio (SINR) is eventually saturated due to increasing interference power. 
The use of a large number of BS antennas exceeding the number of users per cell can effectively reduce interference. SINR is shown to increase with the number of antennas at BS for a single cell with multiple users communicating on the same channel\cite{Mazetta06}. Different types of impairments, e.g., channel estimation errors and intra-cell interference, are diminished with the number of BS antennas. It makes massive multi-input multi-output (MIMO) system promising for future cellular networks. Prototypes of massive MIMO systems have been demonstrated in\cite{Erran12,Young13}.

However, deployment of massive MIMO systems for a multi-cell environment is faced with a non-vanishing inter-cell interference (ICI), called {\em pilot contamination}, resulted from unavoidable reuse of training sequences across cells\cite{Mazetta10}. Mitigation and analysis for pilot contamination have been extensively studied in \cite{Jose11,Hoy11,Huh11,AsMa12}. While pilot contamination was found by using conjugate beamforming in \cite{Mazetta06}, Reference \cite{Jose11} considered a zero-forcing (ZF) and minimum mean square estimation (MMSE) based downlink precoding to mitigate pilot contamination. In \cite{Hoy11}, the impact of pilot contamination was asymptotically analyzed by modeling antenna correlation and as a parameter of the ratio between the number of users per cell and that of BS antennas. To avoid pilot contamination, BS clustering and scheduling were studied in \cite{Huh11}. Part I of the paper\cite{Alex14} and \cite{AsMa12} proposed the use of multi-cell precoding to cancel the pilot contamination. Analysis shows that with an unlimited number of antennas at BSs, the received SINR at each user recovers the linear growth with the number of antennas. The precoding requires exchanging user symbols and large-scale fading coefficients among BSs.

Most of the aforementioned papers, e.g.,\cite{Jose11,Hoy11,AsMa12}, focus on the scenario with an unlimited number of antennas either to simplify analysis or to study an asymptotic scenario. To consider a practical operation regime, this paper focuses on massive MIMO systems with a finite large number of antennas. Part I \cite{Alex14} has shown that, besides pilot contamination, other types of interference become prominent. The proposed precoding based on a ZF design cannot achieve good performance on the $5\%$ outage rates of all users. A balance among these interference, instead of completely eliminating the pilot contamination, turns out to be critical for the designs in this operation regime. We explore the capability of massive MIMO to provide equal quality of service (QoS) for all users in the network by formulating the precoding designs as an optimization problem to maximize the minimum received SINR of all users across the entire network.

Multi-cell cooperation has also been studied in \cite{ZaHanly12} for massive MIMO to provide equal QoS. Our paper differs from \cite{ZaHanly12} in three aspects. First, they considered two cells, whereas we consider an arbitrary number of cells. Secondly, their proposed formulation is to minimize transmit power with SINR constraints at each user. We consider maximization of minimum SINR with individual BS power constraints. Finally, the proposed cooperation in \cite{ZaHanly12} is based on the small-scale fading coefficients. Since the small-scale fading coefficients usually change rapidly, different levels of cooperation (single-cell processing, coordinated beamforming, and macroscopic beamforming) were discussed in \cite{ZaHanly12} to reduce the amount of information exchange. We also aim at reducing the cost of information exchange by precoding design based on the {\em large-scale fading coefficients}. The large-scale fading coefficients depend on the path-loss model and are independent of a specific antenna on one BS. Then, the channel vector between one BS and one user is represented by only one large-scale fading coefficient. Thus, passing large-scale fading coefficients requires less bandwidth on backhaul link compared to small-scale fading coefficients. Further, since the large-scale fading coefficients change more slowly than the small-scale fading coefficients, designs using the large-scale fading coefficients is potentially robust to user mobility.


Our paper takes an optimization approach for precoding designs. There is
a large body of papers formulating problems in communication networks as resource optimization. The classical problem on joint beamforming and power allocation (PA) designs in a single-cell downlink network is particularly related. Several formulations have been considered including sum-rate maximization under a sum-power constraint\cite{StHa06}, power minimization under SINR constraints\cite{RaLiu98,WiSh06,YuLan07}, and minimum-SINR maximization under a sum-power constraint (max-min SINR)\cite{StHa06,WiSh06,Chiang11}. Among these optimization formulations, the power minimization formulation can be translated to a convex optimization and efficiently solved\cite{YuLan07}. This approach was applied for massive MIMO systems in \cite{ZaHanly12}. The max-min SINR formulation was proved to be quasi-convex in \cite{StHa06}, and studied in \cite{Chiang11} through the uplink-downlink duality. The idea of uplink-downlink duality \cite{RaLiu98} is a frequently used tool to reduce the optimization complexity. The uplink-downlink duality discovers the achievable SINR region for an uplink system with received beamforming and transmit PA to be the same as that for a downlink system with transmit beamforming and PA, when both systems are under the same sum-power constraint. This allows coupled downlink beamformers to be computable from uplink beamformers, where they are decoupled and analytically solvable. Yu \cite{YuLan07} connected the uplink-downlink duality with the Lagrangian duality.

In this paper, existing optimization techniques in the aforementioned literature have been considered for our study on massive MIMO systems. After we finished the precoding designs, we discovered that the proposed algorithms are in some sense similar to results in \cite{Chiang11}. We would like to point out that the SINR expression in the context of massive MIMO is more complicated and general compared to a single-cell downlink system. Further, our proof of main duality theorem is based on a necessary condition of max-min SINR optimization, and is distinct from \cite{Chiang11}.

The main results in this paper are summarized as follows
\begin{enumerate}
\item To improve the worst-user's transmission rate in the network, we formulate multi-cell precoding designs to maximize the minimum rate with individual BS power constraints. The precoding designs are based on the large-scale fading coefficients that can be tracked and shared among BSs with reasonable resources. We prove that, similar to single-cell downlink system\cite{StHa06}, the max-min formulation is also quasi-convex, thus optimally solvable by the bisection method and feasibility checking. Simulation results show significant improvement on the $5\%$ outage rate compared to existing methods, e.g., the ZF designs\cite{Alex14} and the PA only method.
\item To simplify computation complexity, we relax the individual BS power constraints to a sum-power constraint over all BSs. We envision the future power amplifier would have higher peak to average-power ratio and power efficiency. Thus, only the total power consumption from all BSs matters. We
decompose the precoding design as beamforming and PA. Under the sum-power constraint, we show that uplink-downlink duality still holds for multi-cell massive MIMO. For the max-min formulation, the optimal downlink beamformer is the same as the optimal uplink beamformer in the virtual uplink system. Based on these observations, we propose an efficient algorithm, called \emph{duality algorithm}, that designs downlink precoding by solving the virtual uplink system at first.
\item We further apply the duality algorithm to solve our original problem with individual BS power constraints due to its low complexity. We prove the proposed suboptimal algorithms can achieve at least $\frac{1}{L}$ ($L$ denotes the number of cells) of the optimal performance. Simulation results show that the suboptimal algorithms can achieve more than $80\%$ of the optimal performance.
\end{enumerate}

The paper is organized as follows. Section \ref{sec-model} describes channel model and protocols for massive MIMO systems. In Section \ref{sec-partI}, we summarize main results obtained in Part I\cite{Alex14}. We design the optimal precoding with individual BS power constraints and one sum-power constraint in Sections \ref{sec-OpPCP} and \ref{sec-duality}, respectively. Section \ref{sec-PerBS} presents low-complexity suboptimal algorithms for the formulation with individual BS power constraints. Section \ref{sec-simulation} provides simulation results and conclusions are given in Section \ref{sec-conclusion}.

\textbf{Notations:} We denote the set of real numbers and complex numbers as $\mathds{R}$ and $\mathds{C}$, respectively. For two matrices $\mb{A}$ and $\mb{B}$, we define $\mb{A}\in \mathds{R}^{n\times m}$ and $\mb{B}\in \mathds{C}^{n\times m}$
as drawn from the $n\times m$ complex matrix space and the $n\times m$ real matrix space, respectively.
The notations $\mb{A}^\t$, $\mb{A}^*$,
$\tr(\mb{A})$, $(\mb{A})_{m,n}$, $\vec(\mb{A})$, and $\|\mb{A}\|$ are used for transpose, Hermitian, trace, its $(m,n)$-th entry, column-wise extension, and Frobenius norm, respectively. For a vector $\mb{a}$, we use $\diag \mb{a}$ to denote a diagonal matrix whose $n$-th diagonal entry is the $n$th entry in the vector $\mb{a}$. For a positive matrix $\mb{A}\in \left(\mathds{R}^+\right)^{n\times n}$ with each entry in $\mb{A}$ being positive, we denote $\lambda\{\mb{A}\}$ as its Perron Frobenius eigenvalue. For random variable $a$, we denote $\Exp a$ as its expectation. The notation $\otimes$ is used for matrix Kronecker product. Finally, we use $\mc{CN}(0,\sigma^2)$ and $\mc{N}(0,\sigma^2)$ for a circular symmetric complex Gaussian distribution and real Gaussian distribution, respectively, both with zero mean and variance $\sigma^2$.

\section{System Model}\label{sec-model}
\begin{figure}
\centering
\includegraphics[width=2.2in]{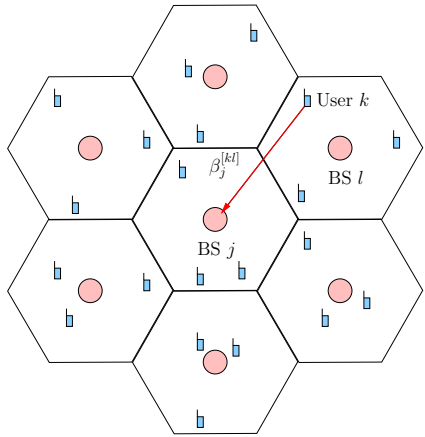}
\caption{Hexagonal cellular model with $L=7$ cells and $K=3$ users per cell. The large-scale fading coefficient between BS $j$ and User $k$ in Cell $l$ is denoted as $\beta_j^{[kl]}$. }\label{fig-chanmodel}
\end{figure}

We consider a two-dimensional hexagonal cellular model with $L$ cells and $K$ users per cell. Each cell is covered by one BS. An example of one layer of hexagonal cells with $L=7$ and $K=3$ is illustrated in Fig.~\ref{fig-chanmodel}. Each BS is equipped with $M$ antennas, and each user has a single antenna. We assume that all antennas are omni-directional.

The channel coefficient (or path) from the $m$-th antenna of the $j$-th BS to the $k$-th user in the $l$-th cell is denoted as ${g}_{mj}^{[kl]}$. We assume that each channel coefficient can be decomposed as a product of the large-scale fading coefficient and the small-scale fading coefficient,
   \begin{align}{g}_{mj}^{[kl]}=\sqrt{\beta_j^{[kl]}}h_{mj}^{[kl]}.
   \end{align}
We denote the {\em channel vector} from the $j$-th base station  to the $k$-th user in the $l$-th cell by
${\mb{g}}_j^{[kl]}=\left[{g}_{1j}^{[kl]},\ldots, {g}_{Mj}^{[kl]}\right]^{\t}$.

The small-scale fading coefficients $h_{mj}^{[kl]}$ are modeled as Rayleigh fading, i.e., an i.~i.~d. $\mc{CN}(0,1)$ distributed random variables. We assume that there is no multipath interference as it can be efficiently eliminated by orthogonal frequency division multiplex (OFDM) signaling. In what follows, we consider only one OFDM tone and therefore we omit tone indices. We assume the block fading model for the small-scale fading coefficients, i.e., they stay constant during {\em small-scale coherent interval} of $T$ OFDM symbols. Within any small-scale coherence interval, the small-scale fading coefficients are assumed being independent from the ones in other small-scale coherence intervals.

The large-scale fading coefficient $\beta_j^{[kl]}\in \mathds{R}^{+}$ depends on the distance between the user and BS and shadowing.
Typically, the distance between BS and a user is significantly larger than the distance between antennas, then the large-scale fading coefficient does not depend on antenna index.
For large-scale fading coefficients, we also assume block fading model, i.e., these coefficients stay constant during {\em large-scale coherence interval}  and are independent from ones in other large-scale coherence intervals. The large-scale coherence intervals are typically longer than the small-scale coherence intervals. For a mobile terminal, the small-scale coherence intervals are constant within $\frac{1}{4}$ of carrier wavelength, whereas the large-scale coherence intervals can stay unchanged for a distance of $10$ wavelengths\cite{dtse}. Therefore, we assume that the large-scale fading coefficients can be accurately estimated and tracked.

\begin{figure}
\centering
\includegraphics[width=2.6in]{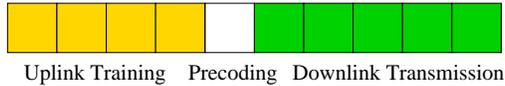}
\caption{TDD protocol in the case of small-scale coherence interval of $T=10$ OFDM symbols:
$\tau=4$ OFDM symbols for uplink training, the duration of $1$  OFDM symbol for processing precoding, and $5$ OFDM symbols for downlink transmission.}\label{fig-TDD}
\end{figure}

We assume that the following time division duplex(TDD) protocol \cite{Mazetta06} is used. The protocol allows  BS to estimate channel coefficients and form beamforming vectors for the corresponding users. Fig.~\ref{fig-TDD} illustrates the protocol. Let $s\kj$ be the symbols intended for transmission to the $k$-th user located in the $j$-th cell. A description of the protocol is provided as follows. A more detailed description and analysis of this protocol can be found in
Part I of the paper\cite{Alex14}.

\noindent{\bf TDD Protocol}
\begin{enumerate}
\item During the first $\tau$ OFDM symbols of one small-scale coherence interval, all users synchronously send their
training sequences $\br^{[k]},\;k=1,\ldots,K$,  and
the $j$-th BS receives $M\times \tau$ matrix
$$
\bY_j=\sqrt{\rho_r\tau } \sum_{k=1}^K\sum_{l=1}^L \bg_j^{[kl]} \br^{[k]^\dag} + \bW_j,
$$
where $\rho_r$ denotes reverse link(uplink) transmit power from each user and $\bW_j\in \mathbb{C}^{M\times \tau}$ is the additive white Gaussian noise matrix with i.i.d. ${\cal CN}(0,1)$ entries.
\item Then, using a duration of $1$ OFDM symbol, the $j$-th BS computes
the MMSE estimates
$$
\hat{\bg}_j\kj={\sqrt{\rho_r\tau}\beta_j^{[kj]}\over 1+\sum_{s=1}^L \rho_r\tau \beta_j^{[ks]}}
\bY_j\br^{[k]}
$$
of the channel vectors $\bg_j\kj$.
\item Finally, during the last $T-\tau-1$ OFDM symbols of one small-scale coherence interval, the
$j$-th BS uses $\hat{\bg}_j\kj,\;k=1,\ldots,K$, as beamforming vectors for small-scale precoding, i.e., it transmits from its $M$ antennas the vector
\begin{align}\label{eq-SmallPre}
\bx_j=\sqrt{\rho_f}
\sum_{k=1}^K {\hat{\bg}_j^{[kj]^\dag} \over \left\|\hat{\bg}_j^{[kj]^\dag}\right\| }  s\kj,
\end{align}
where $\rho_f$ denotes the forward link (downlink) BS transmit power. 
\end{enumerate}
\noindent{\bf The End}

Note that in Step 1 of the TDD protocol, we assume that the same orthonormal training sequences $\br^{[k]},\;k=1,\ldots, K$, are assigned to $K$ users in each cell and that each training sequence is a $\tau$-tuple.

\section{Part I Results Summary}\label{sec-partI}
The use of orthogonal training sequences in the TDD protocol allows a BS to reduce, but not completely eliminate, intra-cell interference. Strong inter-cell interference is still present. It is shown in Part I of the paper\cite{Alex14} that the combined inter-cell and intra-cell interference has several components. One of them, caused by the pilot-contamination effect, is called {\em directed interference}. While other sources of interference and additive noise are vanishing  as $M$ grows, the directed interference grows together with $M$. Thus, for sufficiently large $M$, it becomes the main source of interference. To mitigate the interference, we proposed in Part I a multi-cell precoding scheme called {\em Large-Scale Fading Precoding} (LSFP).
When the number of base station antennas $M$ is large, the LSFP provides an SINR linearly growing with $M$ at each user's receiver.

LSFP is based on limited cooperation between BSs and a network controller. To make this cooperation feasible, we assume that
\begin{itemize}
\item large-scale fading coefficients $\beta_j\kl,\;k=1,\ldots,K,\;j,l\in 1,\ldots,L$, can be accurately estimated and are accessible to the network controller;
\item data signals $s\kl,\;k=1,\ldots,K,\;l=1,\ldots,L$, that are intended for transmission to all users in the network are accessible to all BSs in the network.
\end{itemize}
A formal description of LSFP can be presented as follows.

\noindent{\bf Large-Scale Fading Precoding (LSFP)}
\begin{enumerate}
\item  In the beginning of each large-scale coherence interval, the $j$-th BS estimates coefficients
$\beta_j^{[kl]},\;k=1,\ldots, K,\;l=1,\ldots, L$, and sends them to the network controller.

\item For each large-scale coherence interval,
the network controller computes $L\times L$ LSFP precoding matrices
$$\mb{\Phi}^{[k]}=\left(\begin{array}{c}
\underline{\phi}_1^{[k]}\\
\underline{\phi}_2^{[k]}\\
\vdots\\
\underline{\phi}_L^{[k]}
\end{array}\right),\; k=1,\ldots,K,
$$
as a function of $\beta_j^{[kl]},\;j,l=1,\ldots,L$. It further sends
the $j$-th rows $\underline{\phi}_j^{[k]},\;k=1,\ldots,K$, to the $j$-th BS.

\item The $j$-th BS follows the TDD protocol, but at Step 3 of the TDD protocol  it uses the signal
\begin{equation}\label{eq:tilde s}
{c}_j^{[k]}= \underline{\phi}_j^{[k]} \left(\begin{array}{c}
s^{[k1]}\\
s^{[k2]}\\
\vdots\\
s^{[kL]}
\end{array}\right),\;j=1,\ldots,L,\;k=1,\ldots,K,
\end{equation}
instead of the downlink signal $s\kj$.
\end{enumerate}
\noindent{\bf The End of LSFP}
\vspace{0.1cm}

\begin{figure}[htb]
\centering
\includegraphics[scale=0.7]{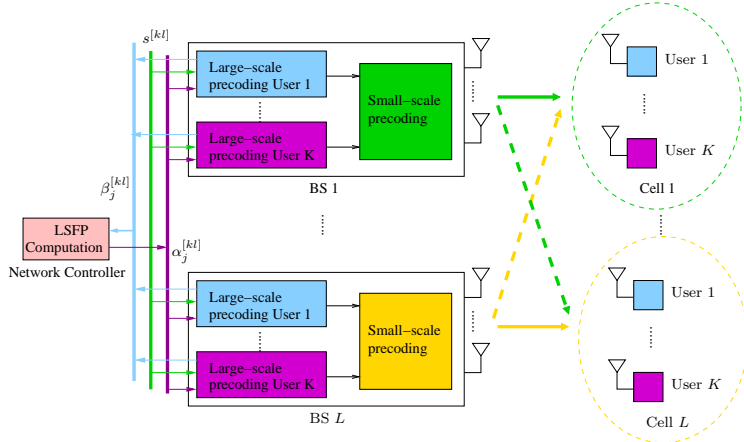}
\caption{System diagram for LSFP. }
\label{fig:BlockDiag}
\end{figure}

The block diagram of this protocol is shown in Fig.\ref{fig:BlockDiag}. One can see that each BS performs two precodings. First, it conducts LSFP (multi-cell) precoding according to (\ref{eq:tilde s}). This precoding is based on the large-scale fading coefficints. Next, in Step 3 of the TDD protocol, a BS performs local precoding according to \eqref{eq-SmallPre}. This precoding is based on the estimated small-scale fading coefficients and is conducted completely locally, that is, it does not require any communication between the BS and the network controller.

The following theorem gives a lower bound on the performance of the LSFP protocol\cite{Alex14}.

\begin{theorem} \label{thm:SINRkl}
The average transmit power of the $j$-th BS is
\begin{align}\label{eq:power_constr_perBS}
\gamma_j={\mathbb E}
[\left\|\mathbf{x}_j\right\|^2]=M \sum_{k=1}^K\left(1+\rho_r\tau\sum_{s=1}^L\beta_j^{[ks]}\right)
\left(\sum_{v=1}^L |{\alpha}_j^{[kv]}|^2\right), j=1,\ldots,L.
\end{align}
Assuming that $\gamma_j\le 1,j=1,\ldots,L$, the downlink transmission rate to the $k$-th user in the $l$-th cell, is lower-bounded by
\begin{align*}
R\kl\ge \log_2(1+\mr{SINR}\kl),
\end{align*}
where the SINR can be expanded as
\begin{align}\label{eq-SINR00}
\mr{SINR}^{[kl]}=\frac{MJ_0^{[kl]}}{\frac{1}{M}+MJ_1^{[kl]}+J_2^{[kl]}},\;k=1,\ldots,K;\;l=1,\ldots,L,
\end{align}
with
\begin{align}
&J_0^{[kl]}=\rho_f\rho_r\tau \left|\sum_{j=1}^L \beta_j\kl{\alpha}_j^{[kl]}\right|^2,~
J_1^{[kl]}=\rho_f\rho_r\tau \sum_{v=1\atop v\not=l}^L \left|\sum_{j=1}^L \beta_j\kl  {\alpha}_j^{[kv]}\right|^2,\nonumber\\
&J_2^{[kl]}=\rho_f\sum_{j=1}^L \sum_{n=1}^K \beta_j\kl(1+\sum_{s=1}^L
\rho_r\tau \beta_j^{[ns]})\left(\sum_{v=1}^L \left|{\alpha}_j^{[nv]}\right|^2\right),
{\alpha}_j^{[kl]} = {\sqrt{\rho_r\tau}\beta_j\kj\over 1+\sum_{s=1}^L \rho_r \tau \beta_j\ks} \phi_j\kl, \nonumber
\end{align}
where $\phi_j\kl$ denotes the $l$-th entry in LSFP vector $\underline{\phi}_j^{[k]}$.
\end{theorem}
Note that the coefficients ${\alpha}_j\kl$ are in one-to-one mapping with the LSFP coefficients ${\phi}_j\kl$. In what follows, we will work with ${\alpha}_j\kl$, since this allows us to shorten notations.

The interference term $MJ_1\kl$ is caused by the pilot contamination effect and the term $J_2\kl$ is a sum of three other sources of interference and additive noise. The term $MJ_1\kl$ grows together with $M$. Hence, in the case of very large $M$ this term becomes the main source of interference. If we use Zero-Forcing LSFP (ZF-LSFP), i.e., we choose ${\alpha}_j\kl,j=1,\ldots,L$, such that
\begin{equation}\label{eq:ZF-PCP}
\sumjL {\alpha}_j\kl \beta_j\kl=0
\end{equation}
we completely eliminate the term $MJ_1\kl$. In this case, we have
$\underset{M\rightarrow \infty}{\lim} \mr{SINR}\kl=\infty$. However, in the regime of finite $M$, e.g., $M=100$, ZF-LSFP has low $5\%$ outage rate, as shown in Part I of the paper\cite{Alex14}. This is because it does not take into account all sources of interference.

\section{Optimal Precoding for a Finite Number of Antennas}\label{sec-OpPCP}
This section focuses on the optimal LSFP design based on the SINR expression in Theorem \ref{thm:SINRkl} with a finite number of antennas at each BS. With finite $M$, the impact of $J_2\kl$ in \eqref{eq-SINR00} can not be ignored. A simple analysis is as follows. Assume that we do not use LSFP, i.e., $\hat{\alpha}_j\kl=c\delta_{j,l}$, where $c$ denotes a constant to satisfy power constraints $\gamma_j\le 1$, the value of ${J}_2^{[kl]}$ can be lowerbounded by
$$J_2\kl \ge c^2\rho_f\rho_r\tau
\sum_{j=1}^L\beta_j^{[kl]}\sum_{n=1}^K\beta_j^{[nj]}, \mbox{ and }
 J_1^{[kl]}=c^2\rho_f\rho_r\tau \sum_{v=1, v\neq l}^L\left|\beta_v^{[kl]}\right|^2.
 $$
Note that the lowerbound of $J_2^{[kl]}$ contains the sum of $\beta_j^{[nj]}$ over both $n$ and $j$, while $J_1^{[kl]}$ has the sum of $\beta_v^{[kl]}$ between User $k$ in Cell $l$ and BSs from other cells. Typically, coefficients $\beta_l\kl$ are larger than coefficients $\beta_v\kl,v\not=l$. Thus, for realistic $M \in [20,1000]$, we have ${J}_2^{[kl]}>J_1^{[kl]}$.
This indicates that to find the optimal or near optimal LSFP, we have to take into account both $MJ_1\kl$ and $J_2\kl$.

We adopt optimization approach for LSFP designs. In Subsection \ref{subsec-form}, we formulate optimization of the precoding designs. Subsections \ref{subsec-quasiconcave} and \ref{subsec-nec} show mathematical properties of the formulation that help to find the global optimal points with low complexity. 

\subsection{Problem formulation}\label{subsec-form}
A widely used engineering criterion for measuring the throughput of multi-user wireless communication systems
is to consider transmission rates $R\kl$ (defined in Theorem \ref{thm:SINRkl})  as random variables and use
{\em $5\%$-outage rate $R_{\mr{out}}$} defined by
\begin{equation}\label{eq-rate}
\Prob\left(R^{[kl]} \le R_{\mr{out}}\right)=0.05.
\end{equation}
By the expression of $R\kl$, the randomness is brought by the large-scale fading coefficients rather than the typically used small-scale fading coefficients. For massive MIMO systems, small-scale fading coefficients are averaged out with the large number of antennas at the BS. Thus, the large-scale fading coefficients have a more dominant effect on stochastic characteristics. 

The $5\%$-outage rate criterion, however, typically pushes one toward complex optimization procedures that
do not allow obtaining insights on the behavior of massive MIMOs. For this reason, instead, we consider the problem of maximization of the minimum $R\kl$ across all users and cells. According to Theorem \ref{thm:SINRkl}, this problem can be equivalently formulated in terms of SINRs as follows
\begin{align}
\label{eq-maxminPerBS}
\max_{\mb{A}^{[1]},\ldots, \mb{A}^{[K]}}\quad & \min_{k,l}~\mr{SINR}\kl(\mb{A}^{[1]},\ldots,\mb{A}^{[K]})={MJ_0\kl\over
{\frac{1}{M}}+MJ_1\kl + J_2\kl}\\
\mbox{s.t. } & \gamma_j=M\sum_{k=1}^K (1+\rho_r\tau \sum_{s=1:L} \beta_j^{[ks]})\sumvL \left|\alpha_j^{[kv]}\right|^2  \le 1,\;j=1,\ldots,L. \label{eq-PerBSconst}
\end{align}

\begin{remark} Though we do not optimize $R_\mr{out}$ directly, by solving the above max-min problem, we obtain LSFP matrices  $\mb{A}^{[k]}$ that drastically
improve $R_\mr{out}$, as demonstrated in Section \ref{sec-simulation}.

\end{remark}

Note that the numerator and denominator of the objective function \eqref{eq-maxminPerBS} are both second order polynomials. This allows us to rewrite formulation (\ref{eq-maxminPerBS}) in the following matrix form.
Let
\begin{equation}\label{eq:matr A and B}
\mb{A}=\sqrt{M\rho_f\rho_r\tau}\left[\begin{array}{c}
\mb{A}^{[1]} \\ \mb{A}^{[2]}\\ \vdots \\ \mb{A}^{[K]}
\end{array}\right], \mb{B}^{[k]}=\left[\begin{array}{cccc}
\beta_1^{[k1]} & \beta_2^{[k1]} & \ldots & \beta_L^{[k1]}\\
\beta_1^{[k2]} & \beta_2^{[k2]} & \ldots & \beta_L^{[k2]}\\
\vdots & & & \vdots \\
\beta_1^{[kL]} & \beta_2^{[kL]} & \ldots & \beta_L^{[kL]}
\end{array}\right], \mb{B}=\left[\begin{array}{cccc}
\mb{B}^{[1]}&\mb{0}&\cdots& \\\mb{0} & \mb{B}^{[2]}\\\vdots&& \ddots &\mb{0}\\ &&\mb{0}&\mb{B}^{[K]}
\end{array}\right],
\end{equation}
Further define
$$
\hat{b}_{j}^{[kln]}=\beta_{j}^{[kl]}\left(\frac{1}{\rho_r\tau}+\sumsL
\beta_j^{[ns]}\right) \mbox{ and } \tilde{b}_{j}^{[n]}=\frac{1}{\rho_f}\left(\frac{1}{\rho_rT}+\sumsL
\beta_j^{[ns]}\right).
$$
 Thus, ${J}_2^{[kl]}={\sum_{n=1}^L\sum_{j=1}^K} \hat{b}_{j}^{[kln]}\underset{v}{\sum} \left|\alpha_j^{[nv]}\right|^2$. We denote
 $$\hat{\mc{B}}^{[kln]}=\diag \left[\hat{b}_{1}^{[kln]}, \hat{b}_{2}^{[kln]}\ldots, \hat{b}_{L}^{[kln]}\right] \mbox{ and }
 \tilde{\mc{B}}_j=\diag \left[\tilde{b}_{j}^{[1]},\ldots, \tilde{b}_{j}^{[K]}\right]\otimes \mb{1}_j,
 $$
 where $\mb{1}_j$ denotes an $L\times L$ zero matrix with only the $(j,j)$th entry being 1.
Let finally
\begin{align}
\hat{\mc{B}}^{[kl]}=\left[\begin{array}{cccc}
\hat{\mc{B}}^{[kl1]}&\mb{0}&\cdots& \\\mb{0} & \hat{\mc{B}}^{[kl2]}\\\vdots&& \ddots &\mb{0}\\ &&\mb{0}&\hat{\mc{B}}^{[klK]}
\end{array}\right].
\end{align}
Note that $\hat{\mc{B}}^{[kl]}\in \mathds{R}^{KL\times KL}$ and $\tilde{\mc{B}}_j\in \mathds{R}^{KL\times KL}$ are diagonal.

Using these notations, after some calculations, we can rewrite the optimization problem (\ref{eq-maxminPerBS}) using a matrix expression as follows
\begin{align} \label{eq-opt}
\max_{\mb{A}\in \mathds{R}^{KL\times L}} &\min_{k,l} \quad \mr{SINR}\kl(\mb{A})= \frac{|(\mb{B}\mb{A})_{(k-1)L+l,l}|^2}{\frac{1}{M}+\underset{v\neq l}{\sum}|(\mb{B}\mb{A})_{(k-1)L+l,v}|^2+\frac{1}{M}\tr \left(\mb{A}^*\hat{\mc{B}}^{[kl]}\mb{A} \right) },\\
\mbox{s.t.} &\quad \|\tilde{\mc{B}}_j\mb{A}\|\le 1,\quad  j=1,\ldots,L.\label{eq-opt_const}
\end{align}
One difference of the matrix formulation compared to the formulation in \eqref{eq-maxminPerBS} is that all power constraints are independent of $M$ and only the objective function depends on $M$.

\subsection{Quasi-concavity}\label{subsec-quasiconcave}
In this subsection, we prove that the formulation in \eqref{eq-opt} is quasi-concave. The following definitions are needed. Let $\Omega$ be a convex subset of $\mathds{R}^n$, then a function $f(\mb{x}),\mb{x}\in \mathds{R}^n$ is called {\em quasi-convex} if for any $\mb{x},\mb{y}\in \Omega$ and $\lambda\in [0,1]$ we have
\begin{equation}\label{eq:def_q-conv_f}
f(\lambda\mb{x}+(1-\lambda)\mb{y})\le \max \{ f(\mb{x}),f(\mb{y})\}.
\end{equation}
Similarly, a function $f(\mb{x})$ is called {\em quasi-concave} if
\begin{equation}\label{eq:def_q_concave_f}
f(\lambda\mb{x}+(1-\lambda)\mb{y})\ge \min \{ f(\mb{x}),f(\mb{y})\}.
\end{equation}

The following theorem simplifies the finding of an optimal solution of the optimization problem \eqref{eq-opt}.
\begin{theorem}\label{prop4}
 Constraints (\ref{eq-opt_const}) are convex functions and the objective function
\eqref{eq-opt} is quasi-concave.
\end{theorem}
\begin{proof}
The first statement is straightforward since all power constraints are norm of linear transformation of optimization variable $\mb{A}$. Only the second statement needs a proof.

Instead of considering $\mr{SINR}\kl$, we consider the quantity $\Gamma\kl={1\over \mr{SINR}\kl}+1$.
Define $\mathds{R}^{KL\times KL}$ matrices
\begin{equation}\label{eq:X_and_P}
\mb{X}=\mb{B}\mb{A} \mbox { and } {\mb{P}}^{[kl]}=\frac{1}{M}\mb{B}^{-*}\hat{\mc{B}}^{[kl]}\mb{B}^{-1}+\mb{\Phi}^{[kl]},
\end{equation}
where $\mb{\Phi}^{[kl]}\in \mathds{R}^{KL\times KL}$ is a matrix whose only nonzero entry is the $((k-1)L+l,(k-1)L+l)$-th entry equal to $1$. Then, after some manipulations, we have
$$
\Gamma\kl=\frac{\frac{1}{M}+\tr \left({\mb{X}}^*\mb{P}^{[kl]}{\mb{X}} \right)}{\left|{x}_{(k-1)L+l,l}\right|^2},
$$
where ${x}_{(k-1)L+l,l}$ is the $((k-1)L+l,l)$-th entry of ${\mb{X}}$. The value of $\Gamma\kl$ is
invariant if ${x}_{(k-1)L+l,l}$ is multiplied by $-1$. Hence, we can assume that ${x}_{(k-1)L+l,l}>0$.
We can rewrite $\Gamma\kl$ in the form
$$
\Gamma^{[kl]}(\mb{X})=\frac{1}{x_{(k-1)L+l,l}}\left(\frac{1}{Mx_{(k-1)L+l,l}}+f\kl(\mb{X})\right),
$$
where $f\kl(\mb{X})=\frac{\tr \left({\mb{X}}^*\mb{P}^{[kl]}{\mb{X}} \right)}{{x}_{(k-1)L+l,l}}$.
The function $f\kl(\mb{X})$ is convex. To prove this, it is
enough to show that for any $c_1,c_2\in \mathds{R}^{+}$ the following inequality holds
\begin{align*}
c_1\frac{\tr\left(\mb{X}^*\mb{P}^{[kl]}\mb{X}\right)}{x_{(k-1)L+l,l}}+c_2\frac{\tr\left(\mb{Y}^*\mb{P}^{[kl]}\mb{Y}\right)}{y_{(k-1)L+l,l}}\ge \frac{\tr\left((c_1\mb{X}+c_2\mb{Y})^*\mb{P}^{[kl]}(c_1\mb{X}+c_2\mb{Y})\right)}{c_1x_{(k-1)L+l,l}+c_2y_{(k-1)L+l,l}}.
\end{align*}
By multiplying both sides with $c_1x_{(k-1)L+l,l}+c_2y_{(k-1)L+l,l}$, we obtain
\begin{align*}
&c_1\left(c_1+c_2\frac{y_{(k-1)L+l,l}}{x_{(k-1)L+l,l}}\right)\tr\left(\mb{X}^*\mb{P}^{[kl]}\mb{X}\right)+c_2\left(c_2+c_1\frac{x_{(k-1)L+l,l}}{y_{(k-1)L+l,l}}\right)\tr\left(\mb{Y}^*\mb{P}^{[kl]}\mb{Y}\right)\\
&\ge \tr\left((c_1\mb{X}+c_2\mb{Y})^*\mb{P}^{[kl]}(c_1\mb{X}+c_2\mb{Y})\right).\end{align*}
Eliminating all duplicate terms from both sides, we obtain
\begin{align*}
\frac{y_{(k-1)L+l,l}}{x_{(k-1)L+l,l}}\tr\left(\mb{X}^*\mb{P}^{[kl]}\mb{X}\right)+\frac{x_{(k-1)L+l,l}}{y_{(k-1)L+l,l}}\tr\left(\mb{Y}^*\mb{P}^{[kl]}\mb{Y}\right)\ge \tr\left(\mb{X}^*\mb{P}^{[kl]}\mb{Y}\right)+\tr\left(\mb{Y}^*\mb{P}^{[kl]}\mb{X}\right).
\end{align*}
The matrix $\mb{P}^{[kl]}$ is Hermitian, so we can define  vectors $\mb{a}=\sqrt{\frac{y_{(k-1)L+l,l}}{x_{(k-1)L+l,l}}}\vec \left((\mb{P}^{[kl]})^{\frac{1}{2}}\mb{X}\right) $ and $\mb{b}=\sqrt{\frac{x_{(k-1)L+l,l}}{y_{(k-1)L+l,l}}}\vec \left((\mb{P}^{[kl]})^{\frac{1}{2}}\mb{Y}\right) $. This allows us to replace the above inequality with $\mb{a}^*\mb{a}+\mb{b}^*\mb{b}\ge  \mb{a}^*\mb{b}+\mb{b}^*\mb{a}$, which is equivalent to $||\mb{a}-\mb{b}||^2\ge 0$.

 The sum of two convex functions is again convex. Thus,
  $g(\mb{X})=\frac{1}{Mx_{(k-1)L+l,l}}+f\kl(\mb{X})$ is convex. It is well known that a linear transformation, in our case the transformation $\mb{A}=\mb{B}^{-1}\mb{X}$,  preserves the convexity. Hence, $h(\mb{A})=g(\mb{B}^{-1}\mb{X})$ is convex.

  The quantity ${x}_{(k-1)L+l,l}$ is the inner product of the $l$-th row of $\mb{B}^{[k]}$ and the $l$-th column of $\mb{A}^{[k]}$, in other words, it is a linear function of entries of $\mb{A}$.
 It is well known, e.g. see \cite{Boyd}, that the ratio of a convex function and a linear function is a quasi-convex function. Hence $\Gamma\kl$, considered as a function of the entries of $\mb{A}$, is a quasi-convex function.

 From the definition of $\Gamma\kl$, we have $\mr{SINR}\kl={1\over \Gamma\kl-1}$.
 Since $\mr{SINR}\kl>0$, we have that $\Gamma\kl-1>0$. The function $\Gamma\kl-1$ is quasi-convex. It is straightforward from \eqref{eq:def_q-conv_f} and \eqref{eq:def_q_concave_f} that the reciprocal of
 a positive quasi-convex function is a quasi-concave function. This concludes the proof.
 \end{proof}

There are a number of algorithms for numerical solutions of quasi-concave optimization problems \cite{Di80}\cite{Kiwi01}.
Any of them can be used for solving the problem in \eqref{eq-opt}.
We used the bisection method \cite{Arfken85}.
 The resulting algorithm can be described as follows.

\noindent{\bf Algorithm 1. Optimal LSFP with $L$ Power Constraints.}
\begin{enumerate}
\item Initialize with an infeasible value of the objective function $\gamma_{in}$ and a feasible value $\gamma_{fea}$ of (\ref{eq-opt}).
\item Compute $\gamma=(\gamma_{in}+\gamma_{fea})/2$. And check the feasibility of $\gamma$.
\item Update $\gamma_{in}=\gamma$ if it is infeasible; otherwise $\gamma_{fea}=\gamma$.
\item Repeat steps 3 and 4 until $|\gamma_{in}-\gamma_{fea}|$ is small enough.

\end{enumerate}
\noindent{\bf The End}

We would like to note that the quasi-concavity of \eqref{eq-opt} leads to a relative simple method of checking the feasibility in Step 2. We used the SeDuMin package\cite{sedumi} to check the feasibility of \eqref{eq-opt}.

To estimate the complexity of Algorithm 1, we note that for each feasibility checking, we have to work with $KL^2$ variables (there are $KL^2$ entries in $\mb{X}$) and $(K+1)L$ constraints (check achievable SINRs at $KL$ users plus $L$ power constraints). So even for relatively small parameters, like $L=7,K=15$, Algorithm 1 has very high complexity. We will propose several approaches to reduce the computation complexity. The observation described in the following subsection has important values.

\subsection{Optimal LSFP lead to equal SINRs}\label{subsec-nec}

In this subsection, we analyze the values of $\mr{SINR}\kl$ achievable at an optimal solution of the problem \eqref{eq-maxminPerBS}. 
Define
$$
S_\mr{opt}=\max_{\mb{A}^{[1]},\ldots,\mb{A}^{[K]}} \min_{k,l} \mathrm{SINR}^{[kl]}( \mb{A}^{[1]},\ldots,\mb{A}^{[K]}  ) \mbox{ and }
{\mathcal{A}}=\arg\max_{\mb{A}^{[1]},\ldots,\mb{A}^{[K]}} \min_{k,l} \mathrm{SINR}^{[kl]}(\mb{A}^{[1]},\ldots,\mb{A}^{[K]}). 
$$
Let further  $\mb{A}^{[1]^*},\ldots,\mb{A}^{[K]^*}\in {\mathcal{A}}$ be a solution of the problem \eqref{eq-maxminPerBS}. Finally, define diagonal matrices 
$$
\mb{P}^{[k]}=
\left[\begin{array}{cccc}
\sqrt{p^{[k1]}} & 0 & & \\
0 & \sqrt{p^{[k2]}} & \ldots & 0\\
\vdots& & \ddots & 0\\
 & & 0 & \sqrt{p^{[kL]}}
 \end{array}\right],k=1,\ldots,K, 
$$
with $p^{[kl]}>0$. 

 \begin{theorem} \label{thm:All_SINRs_equal,_perBS}
The optimal matrices $\mb{A}^{[k]^*},k=1,\ldots,K$ achieve equal SINR for all users, i.e., 
$$
\mr{SINR}^{[nv]}(\mb{A}^{[1]^*},\ldots,\mb{A}^{[K]^*})=S_\mr{opt},~\forall n,v.
$$
\end{theorem}
\begin{proof}
We first show that the optimal $\mb{A}^{[k]^*}, \forall k,$ almost surely cannot have any zero columns. From the mapping between $\alpha_j\kl$ and LSFP coefficient $\phi_j\kl$ in Theorem \ref{thm:SINRkl}, the $s$-th column of $\mb{A}^{[r]^*}$ reflects how all BSs cooperate the transmission to the $r$-th user in the $s$-th cell. If it is a zero vector, it follows that $J_0^{[rs]}$ is equal to zero. Thus, the optimal value $S_\mr{opt} = 0$. Clearly, for randomly generated large-scale fading coefficients, the existing methods such as ZF-LSFP can bring $S_\mr{opt}>0$. Thus, the optimal $\mb{A}^{[r]^*}$ almost surely cannot have zero columns for any $r$.

From \eqref{eq-maxminPerBS}, we have 
\begin{equation}\label{eq:Skl}
\mr{SINR}^{[kl]}(\mb{A}^{[1]}\mb{P}^{[1]},\ldots,\mb{A}^{[K]}\mb{P}^{[K]})={p^{[kl]}M J_0^{[kl]}\over {1\over M} + M\hat{J}_1+\hat{J}_2},
\end{equation}
where
$$
\hat{J}_1^{[kl]}=\rho_f\rho_r\tau \sum_{v=1\atop v\not=l}^L p^{[kv]}\left|\sum_{j=1}^L \beta_j\kl  \hat{\alpha}_j^{[kv]}\right|^2,
$$
and
\begin{align*} 
\hat{J}_2^{[kl]}&=\rho_f\sum_{j=1}^L \sum_{n=1}^K \beta_j\kl(1+\sum_{s=1}^L
\rho_r\tau \beta_j^{[ns]})\left(\sum_{v=1}^L p^{[nv]}\left|\hat{\alpha}_j^{[nv]}\right|^2\right)\\
&=\sum_{n=1}^K \sum_{v=1}^L p^{[nv]} \left(\rho_f \sum_{j=1}^L \beta_j\kl(1+\sum_{s=1}^L
\rho_r\tau \beta_j^{[ns]}) \left|\hat{\alpha}_j^{[nv]}\right|^2\right). 
\end{align*}
Since matrices  $\mb{A}^{[k]^*},k=1,\ldots,K$, do not have zero columns, it follows that entries in $\hat{J}_2^{[kl]}$
$$
\rho_f \sum_{j=1}^L \beta_j\kl(1+\sum_{s=1}^L
\rho_r\tau \beta_j^{[ns]}) \left|\hat{\alpha}_j^{[nv]}\right|^2>0, \forall n,v.
$$
Hence 
$$
\mr{SINR}^{[kl]}(\mb{A}^{[1]}\mb{P}^{[1]},\ldots,\mb{A}^{[K]}\mb{P}^{[K]})={p^{[kl]}M J_0^{[kl]}\over {1\over M} + \sum_{n=1}^K\sum_{v=1}^L q^{[nv]}p^{[nb]}},
$$
for some positive $q^{[nv]}>0, \forall n,v$. We prove by contradiction. Let us assume that for some $r$ and $s$ we have 
$$
\mr{SINR}^{[rs]}( \mb{A}^{[1]^*},\ldots,\mb{A}^{[K]^*}  )>S_{opt}.
$$ 
 
Define a matrix $\mb{P}^{[r]}$ by $p^{[rs]}=\epsilon, 0<\epsilon<1$, and $p^{[rv]}=1,\forall v\not =s$.
And further let $\mb{P}^{[k]}=\mb{I}_L,k\not=r$ being an identity matrix with size $L\times L$.  
Then, we have that 
\begin{enumerate} 
\item Using the new LSFP $
 \mb{A}^{[k]^*}\mb{P}^{[k]},\forall k,
 $
satisfies the individual power constraints in  \eqref{eq-PerBSconst}.
\item $\mr{SINR}^{[rs]}(\mb{A}^{[1]^*}\mb{P}^{[1]},\ldots,\mb{A}^{[K]^*}\mb{P}^{[K]})$
 decreases and all other $\mr{SINR}^{[nv]}(\mb{A}^{[1]^*}\mb{P}^{[1]},\ldots,\mb{A}^{[K]^*}\mb{P}^{[K]} ), (n,v)\not=(r,s)$ increase.
\end{enumerate} 
Therefore, we can choose an $\epsilon$ so that 
$$
\mr{SINR}^{[nv]}(\mb{A}^{[1]^*}\mb{\Lambda}^{[1]},\ldots,\mb{A}^{[K]^*}\mb{\Lambda}^{[K]})>S_\mr{opt},\forall n,v,
$$
which contradicts the assumption that  $\mb{A}^{[1]^*},\ldots,\mb{A}^{[K]^*}\in {\mathcal{A}}$. This concludes the proof. 
\end{proof}
Theorem \ref{thm:All_SINRs_equal,_perBS} provides a necessary condition at the optimal values, all SINRs are equal. In the following section, we will apply the theorem to reduce the complexity of optimization.



\section{Optimization with Sum Power Constraint }\label{sec-duality}
In this section, we consider a relaxation of the power constraints of the optimization problem \eqref{eq-maxminPerBS}. We replace individual BS power constraints with a sum-power constraint over all BSs and
propose a low complexity iterative algorithm for solving the resulting optimization problem.

This section consists of the following subsections. Subsection \ref{subsec-summot} explains motivations behind the sum-power constraint and properties of the formulation that can be extended from the formulation with individual BS power constraints. In Subsections \ref{subsec-relsig} and \ref{subsec-condition}, we introduce some notions and present downlink analysis with decomposition of power allocation and beamforming, respectively. We define a virtual uplink system and prove a duality relationship with our considered downlink system in Subsection \ref{subsec-UL}. Finally, Subsection \ref{subsec:duality_alg} presents the duality algorithm.

\subsection{Motivation and properties of the optimization with sum power constraint}\label{subsec-summot}
We replace the set of $L$ BS transmit power constraints with one sum power constraint over all BSs
\begin{align}\label{eq-maxmin0}
\max_{\alpha_j^{[kl]}\in \mathds{R}}\quad & \min_{k,l} \mr{SINR}\kl={MJ_0\kl\over
{\frac{1}{M}}+MJ_1\kl + J_2\kl}\\
\mbox{s.t.}\quad & M\sumjL \sum_{n=1}^K (1+\rho_r \tau \sum_{s=1}^L \beta_j^{[ns]})\sumvL \left|\alpha_j^{[nv]}\right|^2  \le L.\label{eq-sumconst}
\end{align}

There are at least two reasons for considering this relaxation.

First, as it will be shown later in the paper, this problem can be solved with lower complexity. This allows us, see Section \ref{subsec:duality_alg}, to find suboptimal solutions of the problem \eqref{eq-opt} with complexity significantly lower than that of Algorithm 1.

Second, one can think of organizing an massive MIMO system based on (\ref{eq-maxmin0}) and (\ref{eq-sumconst}). In modern wireless systems, a BS typically operates near its transmit amplifier maximum power to cover an area as large as possible and to gain power efficiency. The expected requirements for wireless systems of future generations include very high data transmission rates for all, or almost all, users, and low energy consumption. Hence, as cell sizes are getting smaller, a BS will transmit most of the time with lower power to avoid strong interference to other cells and to save energy. However, sometimes, a high peak-power amplifier will still be required to provide peak-rate transmission to users with large channel attenuations.
 Thus, the individual BS power constraint can be relaxed to the power constraint of all BSs. For example, a typical macro-cell BS transmits with power higher than $40$~dBm. We can formulate a sum-power constraint of $40$~dBm over several neighboring BSs to reduce the total power consumption without violating individual BS power constraint. The optimization problem defined by (\ref{eq-maxmin0}) and (\ref{eq-sumconst}) is an appropriate model for such kind of wireless systems.

The following results show that the problem \eqref{eq-maxmin0} can be transformed to an unconstrained form.
\begin{proposition}\label{prop2}
The max-min optimization defined by \eqref{eq-maxmin0} and \eqref{eq-sumconst} is equivalent to the following unconstrained problem
\begin{align}\label{eq-maxmin2}
\max_{\alpha_j^{[kl]}\in \mathds{R}}\quad \min_{k,l} \quad \mr{SINR}\kl =\frac{MJ_0^{[kl]}}{\tilde{J}_2^{[kl]}+MJ_1^{[kl]}},
\end{align}
where $\tilde{J}_2^{[kl]}$ has been modified from $J_2^{[kl]}$ as
\begin{align}\label{eq-T1}
\tilde{J}_2^{[kl]}=\sum_{j=1}^L \sum_{n=1}^K \left(\frac{\rho_r\tau}{L}+ \rho_f\beta ^{[kl]}_j\right) \left(\frac{1}{\rho_r\tau}+\sum_{s=1:L} \beta_j^{[ns]}\right)\sum_{v=1}^L \left|\alpha_j^{[nv]}\right|^2.
\end{align}
\end{proposition}
\begin{proof}
We first show that the optimal value of \eqref{eq-maxmin2} upperbounds that of \eqref{eq-maxmin0}. The sum-power constraint \eqref{eq-sumconst} can be rewritten as
\begin{align}\label{eq-powerconst}
\frac{\rho_r\tau}{L}\sumjL \sum_{n=1}^K \left(\frac{1}{\rho_r \tau}+ \sum_{s=1}^L \beta_j^{[ns]}\right)\sumvL \left|\alpha_j^{[nv]}\right|^2  \le \frac{1}{M}.
\end{align}
By replacing $\frac{1}{M}$ in \eqref{eq-maxmin0} with the left-hand side of \eqref{eq-powerconst}, we obtain the objective function of \eqref{eq-maxmin2}. Clearly, for the same $\alpha_j^{[kl]}$, \eqref{eq-maxmin2} upperbounds \eqref{eq-maxmin0}. Additionally, since \eqref{eq-maxmin2} is unconstrained, the optimized value of \eqref{eq-maxmin2} upperbounds that of \eqref{eq-maxmin0}.

Next, we show that the optimized value of \eqref{eq-maxmin2} is achievable by that of \eqref{eq-maxmin0}. The objective function in \eqref{eq-maxmin2} is scale-invariant in $\alpha_j^{[kl]}$. Hence, if $\alpha_j^{[kl]}$ maximize \eqref{eq-maxmin2} then $c\alpha_j^{[kl]}$ also maximize \eqref{eq-maxmin2}. Choosing  $c$ such that $c\alpha_j^{[kl]}$ achieves the equality in \eqref{eq-maxmin0}, we get the same optimal values for both problems.
\end{proof}

We define the matrix $\mb{A}$ in the same way as in \eqref{eq:matr A and B}, and consider
$\mr{SINR}^{[nv]}$ as a function of $\mb{A}$. Using arguments similar to the ones used in Theorems \ref{prop4} and \ref{thm:All_SINRs_equal,_perBS}, we obtain the following results.
\begin{theorem}\label{thm: SumPower_problem_is_q-convex}
\begin{enumerate}
\item The optimization problem \eqref{eq-maxmin2} is quasi-concave.
\item The optimal $\mb{A}^*$ has equal SINRs of all users.
\end{enumerate}
\end{theorem}

Since \eqref{eq-maxmin2} is quasi-concave, we again can use the bisection method for solving it.

\noindent{\bf Algorithm 2. Optimal LSFP with the Sum Power Constraint.}
\begin{enumerate}
\item
 Run the bisection algorithm defined in Algorithm 1 for the quasi-concave optimization problem \eqref{eq-maxmin2}.
\end{enumerate}
\noindent{\bf The End}

The complexity of Algorithm 2 is slightly smaller than the complexity of Algorithm 1, because less power constraints are used for each feasibility checking. However, the complexity still depends on the efficiency of optimization solver and is very high. In the subsequent of this section, we will present an algorithm with low computation complexity to solve the unconstrained formulation in \eqref{eq-maxmin2}. 

\subsection{The feasible range of relative signal power}\label{subsec-relsig}
Instead of directly finding the max min value in \eqref{eq-maxmin2}, we consider the feasible range of the objective function $\frac{MJ_0^{[kl]}}{\tilde{J}_2^{[kl]}+MJ_1^{[kl]}}$ for all $k,l$, since this allows us to obtain more general results. We first, again, rewrite the objective function using a matrix representation for compactness. We reuse matrices $\mb{A}$ and $\mb{B}$ as defined in \eqref{eq:matr A and B}. Further, define
$b_{j}^{[kln]}=\left(\frac{\rho_f\tau}{L}+\rho_f\beta_{j}^{[kl]}\right)\left(\frac{1}{\rho_rT}
+\underset{s}{\sum}\beta_j^{[ns]}\right)$. Thus, $\tilde{J}_2^{[kl]}=\underset{n,j}{\sum} b_{j}^{[kln]}\underset{v}{\sum} \left|\alpha_j^{[nv]}\right|^2$.  Denote also
 $\mc{B}^{[kln]}=\diag [b_{1}^{[kln]},\ldots, b_{L}^{[kln]}]$, and
\begin{equation}\label{eq-notation}
 \mc{B}^{[kl]}=\left[\begin{array}{cccc}
\mc{B}^{[kl1]}&\mb{0}&\cdots& \\\mb{0} & \mc{B}^{[kl2]}\\\vdots&& \ddots &\mb{0}\\ &&\mb{0}&\mc{B}^{[klK]}
\end{array}\right],
\end{equation}
where $\mc{B}^{[kl]}\in \mathds{R}^{KL\times KL}$ is diagonal. Then, the objective function in \eqref{eq-maxmin2} can be rewritten as
\begin{align}\label{eq-SINR2}
\mr{SINR}\kl=\frac{MJ_0^{[kl]}}{\tilde{J}_2^{[kl]}+MJ_1^{[kl]}}=\frac{|(\mb{B}\mb{A})_{(k-1)L+l,l}|^2}{\frac{1}{M}\tr \left(\mb{A}^*\mc{B}^{[kl]}\mb{A} \right) +\underset{v\neq l}{\sum}|(\mb{B}\mb{A})_{(k-1)L+l,v}|^2}.
\end{align}

Define \textit{relative signal power} as
\begin{align}\label{eq-Gamma}
\Gamma^{[kl]}=\frac{\mr{SINR^{[kl]}}}{1+\mr{SINR^{[kl]}}}=\frac{|(\mb{B}\mb{A})_{(k-1)L+l,l}|^2}{\tr \left({\mb{A}}^*\mb{Q}^{[kl]}{\mb{A}} \right)},
\end{align}
where $\mb{Q}^{[kl]}\in \mathds{R}^{KL\times KL}$ and $\mb{Q}^{[kl]}=\frac{1}{M}\mc{B}^{[kl]}+\mb{B}^{*}\mb{\Phi}^{[kl]}\mb{B}$ with $\mb{\Phi}^{[kl]}$ defined after
 \eqref{eq:X_and_P}. In the subsequent of this section, we consider the relative signal power as optimization target instead of SINR, since it has a simpler expression than SINR in \eqref{eq-SINR2}.

In what follows, it will be convenient to represent the matrix $\mb{A}$
 in terms of beamforming vectors $\mb{v}\kl\in \mathbb{R}^{L\times 1}$ and transmit powers $p\kl\in \mathbb{R}^{+}$ as follows
\begin{align}\label{eq:Avp}
\mb{A}=\left[\begin{array}{ccc}\mb{v}^{[11]}\sqrt{p^{[11]}}&\ldots&\mb{v}^{[1L]}\sqrt{p^{[1L]}}
\\\vdots&\ddots&\vdots\\\mb{v}^{[K1]}\sqrt{p^{[K1]}}&\ldots&\mb{v}^{[KL]}\sqrt{p^{[KL]}}
\end{array}\right].
\end{align}
The beamforming vector $\mb{v}^{[kl]}$ defines how $L$ BSs cooperate to assist the transmission to the $k$-th user in the $l$-th cell. The matrix $\mb{Q}^{[kl]}$ in \eqref{eq-Gamma} is block diagonal
with $L\times L$ submatrices. Denote its $n$-th diagonal block as
$\mb{Q}^{[kln]}, n=1,\ldots,K$. After some calculations, we can represent $\Gamma^{[kl]}$ in the following form
\begin{equation}\label{eq-gammakl}
\Gamma^{[kl]}=\frac{|\mb{b}^{[kl]*}\mb{v}^{[kl]}|^2p^{[kl]}}{\sum_{n=1}^K\sumjL
{\mb{v}}^{[nj]*}\mb{Q}^{[kln]}{\mb{v}^{[nj]}}{p}^{[nj]}},
\end{equation}
where $\mb{b}^{[kl]}=\left[\beta^{[kl]}_1,\ldots,
\beta^{[kl]}_L\right]^{\t}$.


We define the feasible range of the relative signal power range $\Omega_{\mr{D}}$ as
\begin{align}\label{eq-region2}
\Omega_{\mr{D}}=\underset{\{p^{[nv]},\mb{v}^{[nv]}: \forall n,v\}}{\bigcup}\left\{\left[\omega^{[11]}, \omega^{[21]},\ldots,\omega^{[KL]} \right]\left| \omega^{[kl]}\le \Gamma^{[kl]} 
,\ \forall k,l \right. \right\},
\end{align}
 The subscript $\mr{D}$ in \eqref{eq-region2} denotes a downlink system.

 In many multi-user communication scenarios, some users may require higher SINRs than others to guarantee certain quality of service (QoS). Let $\mb{u}=(u^{[11]},u^{[21]},\ldots,u^{[KL]})$ with $||\mb{u}||=1$ be defined as
 \begin{align}
 \frac{(\Gamma^{[11]},\Gamma^{[21]},\ldots,\Gamma^{[KL]})}{\|(\Gamma^{[11]},\Gamma^{[21]},\ldots,\Gamma^{[KL]})\|} = \mb{u}.
 \end{align}
 The vector $\mb{u}$ contains the required relations between relative signal powers (and therefore SINRs) of the users. Also, by conditioning on $\mb{u}$, we only need to consider the feasible range of one variable $\|(\Gamma^{[11]},\Gamma^{[21]},\ldots,\Gamma^{[KL]})\|$ instead of a $KL$-dimensional vector. We call $\mb{u}$ a {\em QoS vector}. Then, we can define the {\em conditional feasible range with a given QoS vector} $\mb{u}$ by
  $$
  \Omega_\mr{D}(\mb{u})=[0,\overline{\Gamma}], \mbox{ where } \overline{\Gamma}=\max_{\{\mb{v}^{[nv]},p^{[nv]}:\forall n,v\}} \left\{\Gamma|\Gamma\le {\Gamma\kl\over u\kl} \forall k,l \right\}.
  $$

\subsection{Conditional feasible range with given beamforming vectors}\label{subsec-condition}
  In this subsection, we analyze the scenario with fixed beamforming vector $\mb{v}^{[kl]}$ in \eqref{eq:Avp}. Denote by  $\mb{p}=(p^{[11]},p^{[21]},\ldots,p^{[KL]})\in \mathds{R}^{KL\times 1}$ the power allocation vector.
For a given QoS vector $\mb{u}$ and beamforming vectors $\mb{v}\kl$, we define the {\em conditional feasible range with given $\mb{u}$ and $\mb{v}\kl$} as
$$
\Omega_\mr{D}(\mb{u},\mb{v}^{[11]},\ldots, \mb{v}^{[KL]})=[0,\overline{\Gamma}], \mbox{ where } \overline{\Gamma}=\max_{\mb{p}} \left\{\Gamma|\Gamma\le {\Gamma\kl\over u\kl}, \forall k,l\right\}.
 $$
We would like to estimate how large the conditional feasible range $\Omega_D(\mb{u},\mb{v}^{[11]},\ldots, \mb{v}^{[KL]})$ could be.

  Denote by $\mb{D}, \mb{U}\in \mathbb{R}^{KL\times KL}$
 the diagonal matrices whose $((k-1)L+l)$-th diagonal entries are
$|\mb{b}^{[kl]*}\mb{v}^{[kl]}|^2$ and $u^{[kl]}$, respectively. Let $\mb{F}\in
\mathbb{R}^{KL\times KL}$ be the matrix defined by
\begin{align}\label{eq-Fmatrix}
(\mb{F})_{(k-1)L+l,(n-1)L+j}={\mb{v}}^{[nj]*}\mb{Q}^{[kln]}{\mb{v}^{[nj]}}.
\end{align}

With these notations, the condition $\Gamma \le \Gamma\kl/ u\kl, \forall k,l, $
can be written in the form
\begin{align}\label{eq-eigdec2}
\mb{D}^{-1}\mb{U}\mb{F}\mb{p}\le \frac{1}{\Gamma}\mb{p}.
\end{align}

Since all the diagonals of $\mb{D}$ and $\mb{U}$ are positive and all entries of $\mb{F}$ are positive, $\mb{D}^{-1}\mb{U}\mb{F}$ is a positive matrix. The power vector $\mb{p}$ is clearly an eigenvector of the matrix $\mb{D}^{-1}\mb{U}\mb{F}$ and all entries of $\mb{p}$ have to be nonnegative. Hence,
according to the Perron-Frobenius theorem on nonnegative matrices \cite{matrix}, there is a unique nonnegative eigenvector corresponding to the maximum eigenvalue $\lambda\{\mb{D}^{-1}\mb{U}\mb{F}\}$. Thus, the conditional feasible range is
\begin{align}\label{eq-condregion}
\Omega_\mr{D}(\mb{u},\mb{v}^{[11]},\ldots, \mb{v}^{[KL]})=\left[0, \frac{1}{\lambda\{\mb{D}^{-1}\mb{U}\mb{F}\}}\right].
\end{align}

\subsection{Virtual uplink system}\label{subsec-UL}
In this subsection, we define a virtual uplink system. We prove that this uplink system has the same feasible range of the relative signal power as our downlink system, and that optimal beamforming vectors for these two systems are also the same. Next, we show that finding the optimal
beamforming vectors of the uplink system is a simpler task.
\begin{figure}
\centering
\includegraphics[width=2.2in]{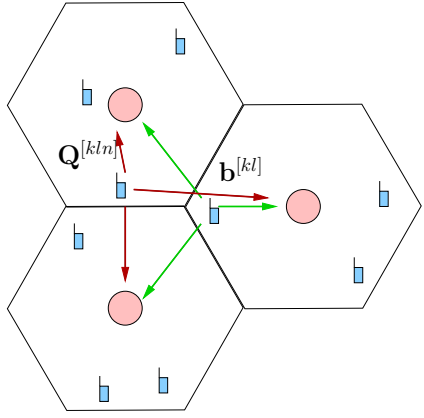}
\caption{A virtual uplink system with $L=3$ and $K=3$ with all BSs connected via backhaul link. The desired channel vector
(green solid lines) for User $k$ in Cell $l$ is modeled as $\mb{b}^{[kl]}$. The covariance matrix of the interference power induced by uplink transmission from User $k$ in Cell $l$ to the reception of User $n$ in all cells (red solid lines) is modeled as
$\mb{Q}_{\mr{U}}^{[kln]}$. User sends uplink signal by power $P_{\mr{U}}^{[kl]}$. All BSs cooperate to decouple User $k$ in Cell $l$'s signal by beamforming vector $\mb{w}^{[kl]}$. Relative signal power is modeled in Eq.~\eqref{eq-gammaUP}.}\label{fig-LSASdual}
\end{figure}

We construct our virtual uplink system as shown in Fig.~\ref{fig-LSASdual}. We denote by $p_\mr{U}\kl$ and $x\kl$ the transmit power and the uplink signal of the $k$-th user in the $l$-th cell.
We assume that all BSs are connected to the network controller (not shown in Fig.~\ref{fig-LSASdual}) and that all coefficients $\beta_j\kl$ are known to the controller. The $j$-th BS receives signal $y_{\mr{U},j}$ and forwards
it to the controller.
The controller computes the estimate of the symbol $x\kl$ as
\begin{align}
\hat{x}\kl=\left[y_{\mr{U},1}, \ldots, y_{\mr{U},L}\right]\mb{w}^{[kl]},
\end{align}
where $\mb{w}^{[kl]}\in
\mathbb{R}^{L\times 1}, \left\|\mb{w}^{[kl]}\right\|=1$, are uplink receive beamforming vectors. The channel paths and interference are modeled using the relative signal power from the $k$-th user in the $l$-th cell as 
\begin{align}\label{eq-gammaUP}
\Gamma^{[kl]}_{\mr{U}}=\frac{p_{\mr{U}}^{[kl]}|\mb{b}^{[kl]*}\mb{w}^{[kl]}|^2}{{\mb{w}^{[kl]*}}\left(\sum_{n=1}^K\sumjL
\mb{Q}_{\mr{U}}^{[njk]}p^{[nj]}_{\mr{U}}\right){\mb{w}^{[kl]}}},
\end{align}
where $\mb{Q}_{\mr{U}}^{[njk]}=\mb{Q}^{[njk]}$ defined in \eqref{eq-gammakl} for the downlink system. For convenience, we will use the downlink $\mb{Q}^{[njk]}$ for the virtual uplink system. Note that though we do not define channel paths and interference explicitly, the above expression is all we need to define the virtual uplink system and to use it for finding solutions for our
downlink system.

The feasible range of the relative signal power in the virtual uplink system can be defined similarly to the downlink system as
\begin{align}\label{eq-uplinkrange}
\Omega_{\mr{U}}=\underset{\{p^{[nv]}_{\mr{U}},\mb{w}^{[nv]}:\forall n,v\}}{\bigcup}\left\{\left[\omega^{[11]}_{\mr{U}}, \omega^{[21]}_{\mr{U}},\ldots,\omega^{[KL]}_{\mr{U}} \right]\left| \omega^{[kl]}_{\mr{U}}\le
\Gamma_\mr{U}\kl, \forall k,l \right. \right\}.
\end{align}

Denote by $\mb{p}_\mr{U}=(p_\mr{U}^{[11]},p_\mr{U}^{[21]},\ldots,p_\mr{U}^{[KL]})\in \mathds{R}^{KL\times 1}$ the power allocation vector.
 Similar to the downlink case, we define {\em QoS vector} $\mb{u}_\mr{U}=(u_\mr{U}^{[11]},u_\mr{U}^{[12]},\ldots,u_\mr{U}^{[KL]})$ and further the {\em conditional feasible range with a given QoS vector} $\mb{u}_\mr{U}$ by
  $$
  \Omega_\mr{U}(\mb{u}_\mr{U})=[0,\overline{\Gamma}], \mbox{ where } \overline{\Gamma}=\max_{\{p_\mr{U}^{[nv]},\mb{w}\nv: \forall n,v\}} \left\{\Gamma|\Gamma\le {\Gamma\kl\over u_\mr{U}\kl}, \forall k,l\right\}.
  $$
   We also define the {\em conditional feasible range given both beamforming vectors and QoS vector} as
$$
\Omega_\mr{U}(\mb{u}_\mr{U},\mb{w}^{[11]},\ldots,\mb{w}^{[KL]})=[0,\overline{\Gamma}], \mbox{ where } \overline{\Gamma}=\max_{\mb{p}_\mr{U}} \left\{\Gamma|\Gamma\le {\Gamma_\mr{U}\kl\over u_\mr{U}\kl}, \forall k,l\right\}.
 $$
For a given $\bu_\mr{U}$ and all beamforming vectors $\mb{w}^{[kl]}$, after computations similar to ones used in \eqref{eq-Fmatrix}, we obtain
\begin{align}\label{eq-eigdecUP}
\mb{D}_{\mr{U}}^{-1}\mb{U}\mb{F}_{\mr{U}}\mb{p}_{\mr{U}}\le \frac{1}{\Omega_{\mr{U}}}\mb{p}_{\mr{U}},
\end{align}
where $\mb{D}_{\mr{U}}$ is an $KL\times KL$ diagonal matrix whose $((k-1)L+l)$-th diagonal entry is $|\mb{b}^{[kl]*}\mb{w}^{[kl]}|^2$, and
\begin{align}\label{eq-FmatrixUP}
(\mb{F}_{\mr{U}})_{(k-1)L+l,(n-1)L+j}={\mb{w}}^{[kl]*}\mb{Q}^{[njk]}{\mb{w}^{[kl]}}.
\end{align}

Thus, with given beamforming vectors $\mb{w}^{[kl]}$,
the conditional feasible range of $\omega_{\mr{U}}$ can be similarly obtained from \eqref{eq-condregion} as

\begin{align}\label{eq-condregionUL}
\Omega_\mr{U}(\mb{u}_\mr{U},\{\mb{w}\kl:\forall k,l\})=\left[0,\frac{1}{\lambda\{\mb{D}_{\mr{U}}^{-1}\mb{U}\mb{F}_{\mr{U}}\}}\right].
\end{align}

The next result establishes the uplink-downlink duality.
\begin{lemma}\label{lemma1}
If we use the same beamforming vectors and the same QoS vectors in the downlink and virtual uplink system, the conditional feasible ranges are equal, i.e.,
$\Omega_\mr{D}(\mb{u}_\mr{U},\mb{w}^{[11]},\ldots,\mb{w}^{[KL]} )=\Omega_\mr{U}(\mb{u}_\mr{U},\mb{w}^{[11]},\ldots,\mb{w}^{[KL]})$.
\end{lemma}
\begin{proof}
If $\mb{v}^{[kl]}=\mb{w}^{[kl]}$, we observe $\mb{F}_{\mr{U}}=\mb{F}^{\t}$ by comparing their definitions in \eqref{eq-FmatrixUP} and \eqref{eq-Fmatrix}, respectively. Similarly, we have $\mb{D}_{\mr{U}}=\mb{D}$. Hence $\mb{D}_{\mr{U}}^{-1}\mb{U}\mb{F}_{\mr{U}}=\mb{D}^{-1}\mb{U}\mb{F}^{\t}$. Taking into account that $\mb{D}^{-1}\mb{U}$ is diagonal, we transform the characteristic polynomial of $\mb{D}^{-1}\mb{U}\mb{F}$ as
\begin{align*}
&\det \left(\mb{D}^{-1}\mb{U}\mb{F}-\lambda \mb{I}\right)=\det \left(\mb{F}\mb{D}^{-1}\mb{U}-\lambda \mb{I}\right)=\det \left(\mb{F}\mb{D}^{-1}\mb{U}-\lambda \mb{I}\right)^{\t}=\det \left(\mb{D}^{-1}\mb{U}\mb{F}^\t-\lambda \mb{I}\right).
\end{align*}
Thus, $\mb{D}^{-1}\mb{U}\mb{F}$ and $\mb{D}^{-1}\mb{U}\mb{F}^{\t}$ have the same maximum eigenvalue $\lambda\{\mb{D}^{-1}\mb{U}\mb{F}\}$. Now,
the accertion follows  from \eqref{eq-condregion} and \eqref{eq-condregionUL}.
\end{proof}
Using this lemma, we have the following results.
\begin{theorem}\label{dualThm}\em{[Duality Theorem]}
\begin{enumerate}
\item For a given QoS vector $\mb{u}_\mr{U}$, we have $\Omega_\mr{D}(\mb{u}_\mr{U})=\Omega_\mr{U}(\mb{u}_\mr{U})$.
 \item Feasible ranges of
the downlink and virtual uplink systems are the same, i.e., $\Omega_\mr{D}=\Omega_\mr{U}$.
\end{enumerate}
\end{theorem}
\begin{proof}
From the definitions of feasible ranges,
it follows that
$$
\Omega_\mr{D}(\mb{u})=\bigcup_{\{\mb{v}\kl:\forall k,l\}} \Omega_\mr{D}(\mb{u},\mb{v}^{[11]},\ldots,\mb{v}^{[KL]}) \mbox{ and } \Omega_\mr{D}(\mb{u}_\mr{U})=\bigcup_{\{\mb{w}\kl:\forall k,l\}} \Omega_\mr{U}(\mb{u},\mb{w}^{[11]},\ldots,\mb{w}^{[KL]}).
$$
Using Lemma \ref{lemma1}, we get
\begin{align*}
&\Omega_\mr{D}(\mb{u})=\bigcup_{\{\mb{v}\kl:\forall k,l\}} \Omega_\mr{D}(\mb{u},\mb{v}^{[11]},\ldots,\mb{v}^{[KL]}) \\
&=
\bigcup_{\{\mb{w}\kl=\mb{v}\kl:\forall k,l\}} \Omega_\mr{U}(\mb{u},\mb{w}^{[11]},\ldots,\mb{w}^{[KL]})=\Omega_\mr{U}(\mb{u}).
\end{align*}
Similarly,  noting that
$\Omega_\mr{D}=\bigcup_{\mb{u}} \Omega_\mr{D}(\mb{u}) \mbox{ and } \Omega_\mr{U}=\bigcup_{\mb{u}_\mr{U}} \Omega_\mr{U}(\mb{u}_\mr{U}),
$
and repeating the same arguments as above,
we got the second claim.
\end{proof}

\subsection{Iterative duality algorithm}\label{subsec:duality_alg}
In this subsection, we present an iterative algorithm for max-min optimization based on duality between downlink and virtual uplink systems.
We rewrite the downlink max-min optimization \eqref{eq-SINR2}  using relative signal power as objective function and $\mb{v}^{[kl]}, \mb{p}$ as  optimization variables,
\begin{equation}\label{eq-DL}
\mr{Downlink}:~S_\mr{D}=\max_{\mb{v}^{[kl]},p^{[kl]}} \min_{k,l}~
\Gamma\kl(\mb{v}^{[11]},\ldots,\mb{v}^{[KL]},\mb{p})=\frac{|\mb{b}^{[kl]*}\mb{v}^{[kl]}|^2 p^{[kl]} }
{\sum_{k=1}^K\sumjL
\mb{v}^{[nj]*}
\mb{Q}^{[kln]}\mb{v}^{[nj]}p^{[nj]}}.
\end{equation}
Similarly, for the virtual uplink system, we formulate a max-min optimization problem as
\begin{equation}\label{eq-UL}
\mr{Uplink}:~ S_\mr{U}=\max_{\mb{w}^{[kl]},p_{\mr{U}}^{[kl]}} \min_{k,l}
\Gamma_\mr{U}\kl(\mb{w}^{[11]},\ldots,\mb{w}^{[KL]},\mb{p}_U)=\frac{p_{\mr{U}}^{[kl]}|\mb{b}^{[kl]*}\mb{w}^{[kl]}|^2}{{\mb{w}^{[kl]*}}
\left(\sum_{n=1}^K\sumjL \mb{Q}^{[njk]}p^{[nj]}_{\mr{U}}\right){\mb{w}^{[kl]}}}.
\end{equation}

In Theorem \ref{thm: SumPower_problem_is_q-convex}, we have proved that there is a  solution of the problem (\ref{eq-DL}) at which $\mr{SINR}^{[nv]}=S_\mr{D}, \forall n,v$. In a similar way, we can prove the following lemma.
\begin{lemma}\label{lemma-nec} There exist beamforming vectors $\mb{w}^{[kl]^*}$ and powers $p_\mr{U}^{[kl]^*}$ such that
$$
\Gamma_\mr{U}\nv(\mb{w}^{[11]^*},\ldots,\mb{w}^{[KL]^*},\mb{p}_\mr{U}^*) = S_\mr{U}, \forall n,v.
$$
\end{lemma}

The next results shows that the beamforming vectors $\mb{v}^{[kl]^*}=\mb{w}^{[kl]^*}, \forall k,l$, give an optimal
solution of \eqref{eq-DL}.
\begin{theorem}\label{thm2}
\begin{enumerate}
\item For the optimization problems in \eqref{eq-DL} and \eqref{eq-UL}, we have $S_\mr{D}=S_\mr{U}$.
\item  There exist powers
$p^{[kl]^*}$ such that
$$
\Gamma_\mr{D}^{[nv]}(\mb{v}^{[11]^*},\ldots, \mb{v}^{[KL]^*},\mb{p}^{*})=S_\mr{D}, \forall n,v.
$$
with $\mb{v}^{[kl]^*}=\mb{w}^{[kl]^*}, \forall k,l$.
\end{enumerate}
\end{theorem}
\begin{proof}
According to Theorem \ref{thm: SumPower_problem_is_q-convex} and Lemma \ref{lemma-nec}, we can look for optimal solutions of \eqref{eq-DL} and \eqref{eq-UL} assuming the QoS vector having equal entries, i.e. $u\kl=u_\mr{U}\kl={1\over \sqrt{KL}}$ for $\forall k,l$. By duality Theorem \ref{dualThm}, we know the feasible range of the relative signal power is the same between uplink and downlink. Thus, along the same QoS vector, virtual uplink and downlink have equal optimal values, i.e., $S_\mr{D}=S_\mr{U}$.

Claim 2 is straightforward by replacing  $\mb{v}^{[kl]^*}=\mb{w}^{[kl]^*}$. This concludes the proof.
\end{proof}
We would like to note that the optimal powers in for the downlink and uplink powers are not the same, i.e.,  $p^{[kl]^*}\neq p_\mr{U}^{[kl]^*}$. Additionally, Theorem \ref{thm2} is based on the relative signal power. Extension to SINR is straightforward.

Following Theorem \ref{thm2}, we provide an algorithm to find downlink beamforming vectors and powers.
First, we compute the optimal beamforming vector and power allocation for the virtual uplink system
assuming $u_\mr{U}\kl={1\over \sqrt{KL}},~\forall k,l$. According to \eqref{eq-UL}, the relative signal power $\Gamma_\mr{U}\kl$ is independent of $\mb{w}^{[nj]}$ for $(k,l)\neq (n,j)$. Hence, for given powers $p_{\mr{U}}^{[kl]}$,  the optimal uplink beamforming vectors can be found by taking the derivative of
\eqref{eq-gammaUP} with respect to $\mb{w}^{[kl]}$ and finding its
zeros. By doing this, we obtain the optimal beamforming vectors
\begin{align}\label{eq-beamformerUL}
\mb{w}^{[kl]}=c\left(\sum_{n=1}^K\sum_{j=1}^L \mb{Q}^{[njk]}p_{\mr{U}}^{[nj]}
\right)^{-1}\mb{b}^{[kl]},\forall k,l.
\end{align}
where $c\in \mathds{R}^+$ is a normalization coefficient to satisfy $\|\mb{w}^{[kl]}\|=1$.
Replacing $\mb{w}^{[kl]}$ in \eqref{eq-gammaUP} with \eqref{eq-beamformerUL}, we obtain
\begin{align}\label{eq-formula1}
\max_{p_{\mr{U}}^{[kl]},\mb{w}^{[kl]}}\quad \min_{k,l}\quad \Gamma^{[kl]}_{\mr{U}} =\max_{p_{\mr{U}}^{[kl]}}\quad \min_{k,l}\quad p_{\mr{U}}^{[kl]}\left(\mb{b}^{[kl]*}\left(\underset{n=1:K,j=1:L}{\sum}\mb{Q}^{[njk]}p_{\mr{U}}^{[nj]}\right)^{-1}\mb{b}^{[kl]}\right).
\end{align}
Note that we assumed that all $u_\mr{U}\kl$ have the same value. From Lemma \ref{lemma-nec}, we have
 $\Gamma_{\mr{U}}^{[kl]}=\Delta_\mr{U},\forall k,l$,  for some
$\Delta_{\mr{U}}\in \mathbb{R}^{+}$.
Thus, we have to solve the following set
of equations
\begin{align}\label{eq-formula2}
p_{\mr{U}}^{[kl]}\left(\mb{b}^{[kl]*}\left(\underset{n=1:K,j=1:L}{\sum}\mb{Q}^{[njk]}p_{\mr{U}}^{[nj]}\right)^{-1}\mb{b}^{[kl]}\right)=\Delta_{\mr{U}},\
\forall k,l.
\end{align}
Note that all $KL$ equations are nonlinear with respect to $p_{\mr{U}}^{[kl]}$.
We propose the following  iterative algorithm to solve \eqref{eq-formula2}.

\noindent{\bf Iterative Power Search Algorithm}
\begin{enumerate}
\item Assign $p_{\mr{U}}^{[kl](0)}=1,\forall k,l$, and repeat several times the following steps
\item $\tilde{p}_{\mr{U}}^{[kl]} = \left(\mb{b}^{[kl]*}\left(\underset{n,j}{\sum}\mb{Q}^{[njk]}p_{\mr{U}}^{[nj](t)}\right)^{-1}\mb{b}^{[kl]}\right)^{-1}$.
\item $p_{\mr{U}}^{[kl](t+1)}=\frac{KL\tilde{p}_{\mr{U}}^{[kl]}}{\underset{k,l}{\sum}\tilde{p}_{\mr{U}}^{[kl]}}$.
\item $t=t+1$ until a fixed number of iterations.
\end{enumerate}
\noindent{\bf The End}

The convergence analysis on our iterative algorithm can be conducted similarly as that on Algorithm 2 from \cite{Chiang11}. The interested readers are referred to their analysis.


The Iterative Power Search Algorithm, \eqref{eq-beamformerUL}, Theorem \ref{thm2},
and \eqref{eq-eigdec2} leads to finding the optimal beamforming vectors  $\mb{v}\kl$ and downlink powers $p\kl$ for the optimization problem \eqref{eq-DL}.
All needed steps are summarized in the following algorithm.

\noindent{\bf Algorithm 3: Uplink-Downlink Duality}\label{Algm-Dual}
\begin{enumerate}
\item Obtain matrices $\mb{Q}^{[njk]}$ defined in \eqref{eq-gammakl}.
\item Compute the uplink powers $p_{\mr{U}}^{[kl]}$ using the Iterative Power Search Algorithm.
\item  Compute the uplink beamforming vectors  $\mb{w}^{[kl]}$ using \eqref{eq-beamformerUL}.
\item Reuse them as the downlink beamforming vectors $\mb{v}^{[kl]}=\mb{w}^{[kl]}, \forall k,l.$.
\item Construct matrices $\mb{D}$ and $\mb{F}$ defined in \eqref{eq-eigdec2}.
\item  Find the optimal downlink powers $p^{[kl]}$ through eigenvalue decomposition of $\mb{D}^{-1}\mb{F}$ using \eqref{eq-eigdec2}.
\end{enumerate}
\noindent{\bf The End of Algorithm 3}

Steps 1,2,3 and 5 in Algorithm 3  involve only multiplication and addition of $KL\times KL$ real matrices.
Step 6 requires one eigenvalue decomposition of a $KL\times KL$ real matrix. The overall complexity of Algorithm 3 is significantly lower than the complexity of Algorithm 2.

\section{Suboptimal Algorithms with Individual BS Power Constraints}\label{sec-PerBS}
In this section, we consider how to use the iterative algorithm (Algorithm 3) proposed for the problem with the sum-power constraint for the optimization \eqref{eq-opt} with per-BS power constraints. Below we propose algorithms that are suboptimal compared to Algorithm 1 defined in Subsection \ref{subsec-quasiconcave}. These algorithms, however, have significantly lower complexity.

We first consider a centralized scenario where a network hub knows all large-scale fading coefficients $\beta_j\kl$ in the entire network and computes $L\times L$ precoding matrices $\mb{A}^{[k]}$. In Subsection \ref{subsec-centalg}, we propose three suboptimal algorithms for the centralized scenario. Subsection \ref{subsec-decenalg} further extends the algorithms to a scenario when there is no centralized node. Instead, each BS knows large-scale fading coefficients of only local neighboring cells and performs its own decentralized algorithm.

\subsection{Centralized suboptimal algorithm}\label{subsec-centalg}
For our main optimization problem \eqref{eq-maxminPerBS} with individual BS power constraints, the duality between the uplink and downlink systems, in general, does not hold. Thus, Algorithm 3 can not be applied directly. Below, we show that Algorithm 3 still can be used to provide low complexity algorithms.

We again assume that the precoding matrix is decoupled into beamforming vectors and transmit powers according to \eqref{eq:Avp}. Based on \eqref{eq-sumconst}, we define the range of precoding coefficients for the sum-power constraint $Z$ as
 \begin{equation}\label{eq-sumcon1}
\Omega_{\mr{sum}}(Z)=\left\{\mb{A}\left| M\sumjL \sum_{n=1}^K (1+\rho_r \tau \sum_{s=1}^L \beta_j^{[ns]})\sumvL \left|\alpha_j^{[nv]}\right|^2 \le Z \right.\right\}.
\end{equation}
When $Z=L$, we have the power constraint \eqref{eq-sumconst}.
 Recall that the power of BS $j$ is defined in \eqref{eq-PerBSconst} as
\begin{align}
\gamma_j(\mb{A})= M\sum_{k=1}^K (1+\rho_r\tau \sum_{s=1:L} \beta_j^{[ks]})\sumvL \left|\alpha_j^{[kv]}\right|^2,
\end{align}
where the function uses the precoding coefficients $\mb{A}$ as variables.

One may try to use
Algorithm 3 with the sum power constraint $Z$ to obtain beamforming vectors $\mb{v}^{[kl]}_{\mr{sum}}$ and powers $p^{[kl]}_{\mr{sum}}$ and to reuse them for the problem \eqref{eq-maxminPerBS}. However, the powers $p^{[kl]}_{\mr{sum}}$ have to be modified to satisfy the per-BS power constraints. That explains the reason that we set $Z$ to be the extra parameter to satisfy individual BS power constraints. A heuristic way is presented below.

\noindent{\bf Algorithm 4}
\begin{enumerate}
\item Run Algorithm 3 with the sum-power constraint $\Omega_{\mr{sum}}(Z)$, where $Z$ can be any number with $1\le Z \le L$. We obtain beamformers $\mb{v}^{[kl]}_{\mr{sum}}$ and powers ${p}^{[kl]}_{\mr{sum}}$.
\item Assign $\mb{v}^{[kl]}_{\mr{sum}}$ as beamforming vectors: $\mb{v}^{[kl]}_{\mr{per}}=\mb{v}^{[kl]}_{\mr{sum}}, \forall k,l$.
\item Compute the assigned powers as ${p}^{[kl]}_{\mr{per}}=
\frac{{p}_\mr{sum}^{[kl]}}{\underset{j}{\max} \gamma_j(\mb{A}_{\mr{sum}})}, \forall k,l$.
\end{enumerate}
\noindent{\bf The End}

Note that $\mb{v}^{[kl]}_{\mr{per}}$ and ${p}^{[kl]}_{\mr{per}}$ are the beamforming vectors and powers used under individual BS power constraints. Step 3 divides ${p}_\mr{sum}^{[kl]}$ by the maximum power used by one BS to satisfy the power constraints.

It is instructive to get lower bounds on the performance of Algorithm 4 with $Z=L$ and $Z=1$. Denote by $\mr{SINR}_{\mr{sum}}^{\mr{o}}$ and $\mr{SINR}_{\mr{per}}^{\mr{o}}$ the optimal
solutions of \eqref{eq-maxminPerBS} and \eqref{eq-maxmin2} respectively.  We first consider the case with $Z=L$.  Since the individual BS power constraint is contained in the sum-BS power constraint, we have  $\mr{SINR}_{\mr{per}}^{\mr{o}}\le \mr{SINR}_{\mr{sum}}^{\mr{o}}$.
Denote further $\tau_1=\underset{j}{\max} \gamma_j(\mb{A}_{\mr{sum}})$, where $\mb{A}_{\mr{sum}}$ is an optimal precoding matrix for the sum-BS power constraint problem. Since $\sum_{j=1}^L \gamma_j(\mb{A}_{\mr{sum}})=L$, we have $\tau_1\ge 1$. Thus, we can bound as
\begin{align}\nonumber
  &\mr{SINR}_{\mr{per}}^{[kl]}=\frac{MJ_0^{[kl]}(\mb{A}_{\mr{sum}})/\tau_1}{\frac{1}{M}+MJ_1^{[kl]}(\mb{A}_{\mr{sum}})/\tau_1+J_2^{[kl]}(\mb{A}_{\mr{sum}})/\tau_1}\\ \nonumber
  &\ge \frac{1}{\tau_1}\frac{MJ_0^{[kl]}(\mb{A}_{\mr{sum}})}{\frac{1}{M}+MJ_1^{[kl]}(\mb{A}_{\mr{sum}})+J_2^{[kl]}(\mb{A}_{\mr{sum}})}= \frac{1}{\tau_1}\mr{SINR}_{\mr{sum}}^{\mr{o}}\\
  &\ge \frac{1}{\tau_1}\mr{SINR}_{\mr{per}}^{\mr{o}}\ge \frac{1}{L}\mr{SINR}_{\mr{per}}^{\mr{o}}, \forall k,l. \label{eq-last}
  \end{align}
The notations $J_i^{[kl]}(\mb{A}_{\mr{sum}})$ are used to denote $J_i^{[kl]}$ that uses $\mb{A}_{\mr{sum}}$ as variables. The first equality holds because $J_0^{[kl]}$, $J_1^{[kl]}$, $J_2^{[kl]}$, and all power constraints are second-order functions of $\mb{A}_{\mr{sum}}$. The first inequality is achieved by upperbounding $J_i^{[kl]}(\mb{A}_{\mr{sum}})/\tau_1 < J_i^{[kl]}(\mb{A}_{\mr{sum}})$ for $i=1,2$. The last inequality in \eqref{eq-last} is achieved because $\tau_1 \le L$. Thus, we have that the achievable SINR of Algorithm 4 with $Z=L$ is within $\frac{1}{L}$ of the optimal value $\mr{SINR}_{\mr{per}}^{\mr{o}}$.

Let now $Z=1$.
Define the optimized SINRs of sum-BS power constraint using constraints $\Omega_{\mr{sum}}(L)$ and $\Omega_{\mr{sum}}(1)$ by $\mr{SINR}_{\mr{sum}}^{\mr{o}}(L)$ and $\mr{SINR}_{\mr{sum}}^{\mr{o}}(1)$, respectively. It can be observed that $\mr{SINR}_{\mr{sum}}^{\mr{o}}(1)\ge \frac{1}{L}\mr{SINR}_{\mr{sum}}^{\mr{o}}(L)$.
Then, we can bound as
$$\mr{SINR}_{\mr{per}}^{[kl]}>\mr{SINR}_{\mr{sum}}^{\mr{o}}(1)\ge \frac{1}{L}\mr{SINR}_{\mr{sum}}^{\mr{o}}(L)\ge \frac{1}{L}\mr{SINR}_{\mr{per}}^{\mr{o}}.
 $$
The first inequality holds because with $Z=1$ in Step 3, we have to increase power of all users by a common factor. Therefore, Algorithm 4 with both $Z=L$ and $Z=1$ can achieve at least $\frac{1}{L}$ of $\mr{SINR}_{\mr{per}}^{\mr{o}}$.

Motivated by these obzervations, we define Algorithm 5, in which we conduct an optimization, over sum power constraint value $Z$. In this algorithm, $\Delta\in \mathds{R}^+$ denotes the step size of line search over power constraint.

  \noindent{\bf Algorithm 5}
 \begin{enumerate}
 \item For {$Z=1:\Delta:L$} do Steps 2.
 \item Run Algorithm 4 with $\Omega_{\mr{sum}}(Z)$.
 \item Choose $Z$ with the best result.
 \end{enumerate}
 \noindent{\bf The End}

Finally, we would like to remark that the proposed suboptimal algorithms can be used to simplify the complexity of the optimal Algorithm 1. We can use Algorithm 4 or 5 to find the initial feasible SINR in the bisection method. Similarly, Algorithm 3 with the sum-power constraint can be used for the initial infeasible SINR. This can potentially reduce the number of iterations of the bisection method.

\subsection{Decentralized suboptimal algorithm}\label{subsec-decenalg}

In all previous sections, we assumed that the network controller is connected to all BSs and get access to all large-scale fading coefficients. We also assumed that any BS has access to all data symbols  intended for transmission to all users across the entire network. This approach can be used for small networks with the number of cells $L$ not exceeding $20$-$30$, like networks covering a small town, campus, or some other dedicated facility.
As $L$ grows, however, this approach becomes unpractical.

In this subsection, we consider a network setting without a centralized controller. Instead, we assume that the $j$-th BS has connections only to the neighboring BSs from a set ${\cal N}(j)$.
 For instance, in the case of hexagonal cells, ${\cal N}(j)$ can be formed by the $j$-th BS itself and its six neighboring BSs. Thus, we assume that the $j$-th BS has access only to the large-scale fading coefficients $\beta_l^{[kr]}$ and data symbols $s^{[kr]}$ for $k=1,\ldots,K$, and $\forall l,r\in {\cal N}(j)$.
 Without losing generality, we assume that ${\cal N}(j)$ has equal size for all BSs. Let $N=|{\cal N}(j)|$ denotes the size of cooperating BSs. We propose the following decentralized algorithm.

\noindent{\bf Algorithm 6. Decentralized LSFP}
\begin{itemize}
\item The $j$-th BS collects $\beta_l^{[kr]}$ and $s^{[kr]}$ for $k=1,\ldots,K$, and $\forall l,r\in {\cal N}(j)$.
\item The $j$-th BS runs any algorithms with individual BS power constraints (e.g.,Algorithm 1, 4, and 5) using these $\beta_l^{[kr]}$ coefficients, and assuming that the entire network consist of Cells $l\in {\cal N}(j)$. As a result, it gets
  $N\times N$ precoding matrices $\mb{A}^{[k]},~k=1,\ldots, K$.
\item The $j$-th BS uses only the row, say $\underline{\alpha}$,
of $\mb{A}^{[k]}$ corresponding to itself and discards all other rows of $\mb{A}^{[k]}$.
\item Finally, the BS computes symbols
$$
c_j^{[k]}=\underline{\alpha} \mb{s}_k^\t, \;k=1,\ldots,K,
$$
where the vector $\mb{s}_k$ is formed by appropriately ordered data symbols $s^{[kr]},r\in {\cal N}(j)$.
\item The $j$-th BS uses $c_j^{[k]}$ at Step 3 of TDD protocol defined in Section \ref{sec-model}.
\end{itemize}
\noindent{\bf The End}

Note that when ${\cal N}(j)$ is equal to the entire cellular network for all $j$, the decentralized algorithm becomes identical to the centralized algorithm.

\section{Simulation}\label{sec-simulation}

This section presents simulated performance of proposed algorithms. We considered the cases of one layer $L=7$ or two layer $L=19$ cells, which are wrapped into a torus \cite{torus}. Wrapping cells into a torus allows us to imitate a network with infinite cells.
The parameters used in simulations are taken according to the 3GPP standard \cite{3GPP}. For each cell, we have $K=10$ users uniformly distributed with an exclusion of central disk with radius $r_\mr{h}=62.5$ meters. Each cell has a radius of one kilometer. We model the large-scale fading coefficient $\beta_j^{[kl]}$ as log-normal distribution based on the Urban Macro model\cite{3GPP}
\begin{align}\label{eq-largescale}
10\log_{10}\beta_j^{[kl]}=-139.5-35\log_{10}d^{[kl]}_j+ \Psi,
\end{align}
where $\Psi$ denotes the shadow fading coefficient with i.~i.~d.~$\mc{N}(0,\sigma^2_{\Psi})$ distribution; and $d^{[kl]}_j$ denotes the distance, measured in kilometer, between the $j$-th BS and the $k$-th user in the $l$-th cell.  The noise variance at each receiver is calculated by $\sigma^2=290\times \kappa \times  B\times \mr{NF}$, where $\kappa$, $B$, and $\mr{NF}$ denote the Boltzmann constant, bandwidth, and noise figure, respectively. We use parameters $B=20$~MHz; noise figures for BS and users are $4$~dB and $9$~dB, respectively. The transmit powers of BS and users are $\rho_f=48$~dBm and $\rho_r=23$~dBm, respectively. For a given LSFP algorithm, we generate random coefficients $\beta_j\kl$ and use the algorithm to compute  data transmission rates $R\kl=\log(1+\mr{SINR}^{[kl]})$. We consider these rates as random variables and plot their empirical cumulative distribution function (CDF).

\begin{figure}[htb]
\includegraphics[scale=0.8]{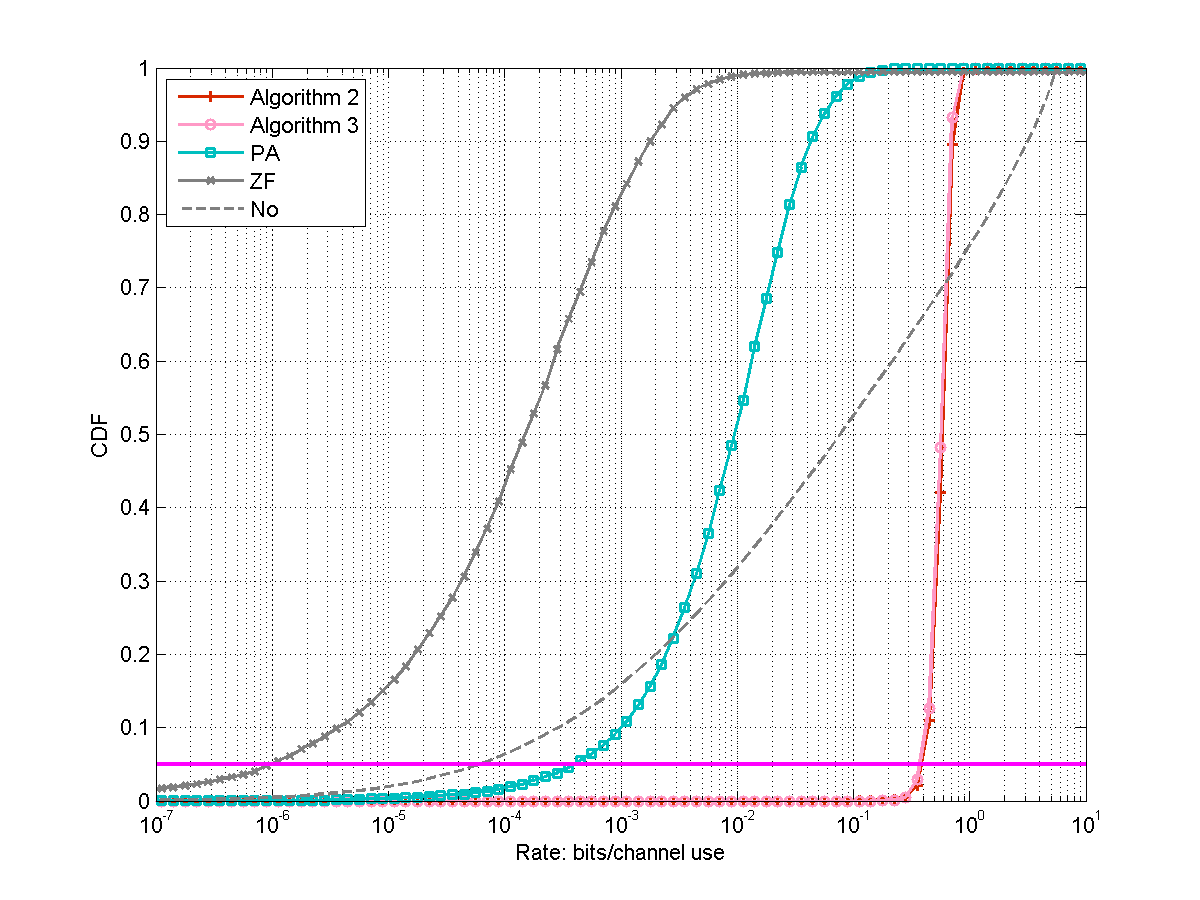}
\caption{The CDF of achievable rates among all users for the proposed algorithms under the sum-power constraint in a cellular network with $L=7$, $K=10$, and $M=64$. The algorithm with only PA but no beamforming is labeled as `PA'; The ZF LSFP is labeled as `ZF'; the algorithm without LSFP is labeled as `No'.}\label{fig-sum_power1}
\end{figure}
In Fig. \ref{fig-sum_power1}, we consider the case of the sum power constraint optimization problem \eqref{eq-maxmin2} with $L=7$ and $M=64$. We observe that Algorithm 2 and Algorithm 3 have identical performance. Note that Algorithm 3 is based on uplink-downlink duality, and has significantly lower complexity compared to Algorithm 2. Further, we see that Algorithms 2 and 3
give dramatic improvement over the case when no LSFP or ZF-LSFP is used. In Fig.~\ref{fig-sum_power1},
we also plot results when optimal PA is used. PA optimizes transmit powers for each user to minimize interference to other users. There exist extensive literature on this subject, e.g.,\cite{Yate95} and references therein. Note that PA corresponds to LSFP with diagonal matrices $\mb{A}^{[k]}$. One can see that at the $5\%$ outage rate defined in \eqref{eq-rate}, Algorithm 2 or 3 achieves $R_{\mr{out}}=0.4$ bits/channel use and the PA algorithm achieves $R_{\mr{out}}=4\times 10^{-4}$ bits/channel use. Thus, LSFP gives very large, about $1000$-fold, improvement in data transmission rates compared with the PA approach.

\begin{figure}[htb]
\includegraphics[scale=0.8]{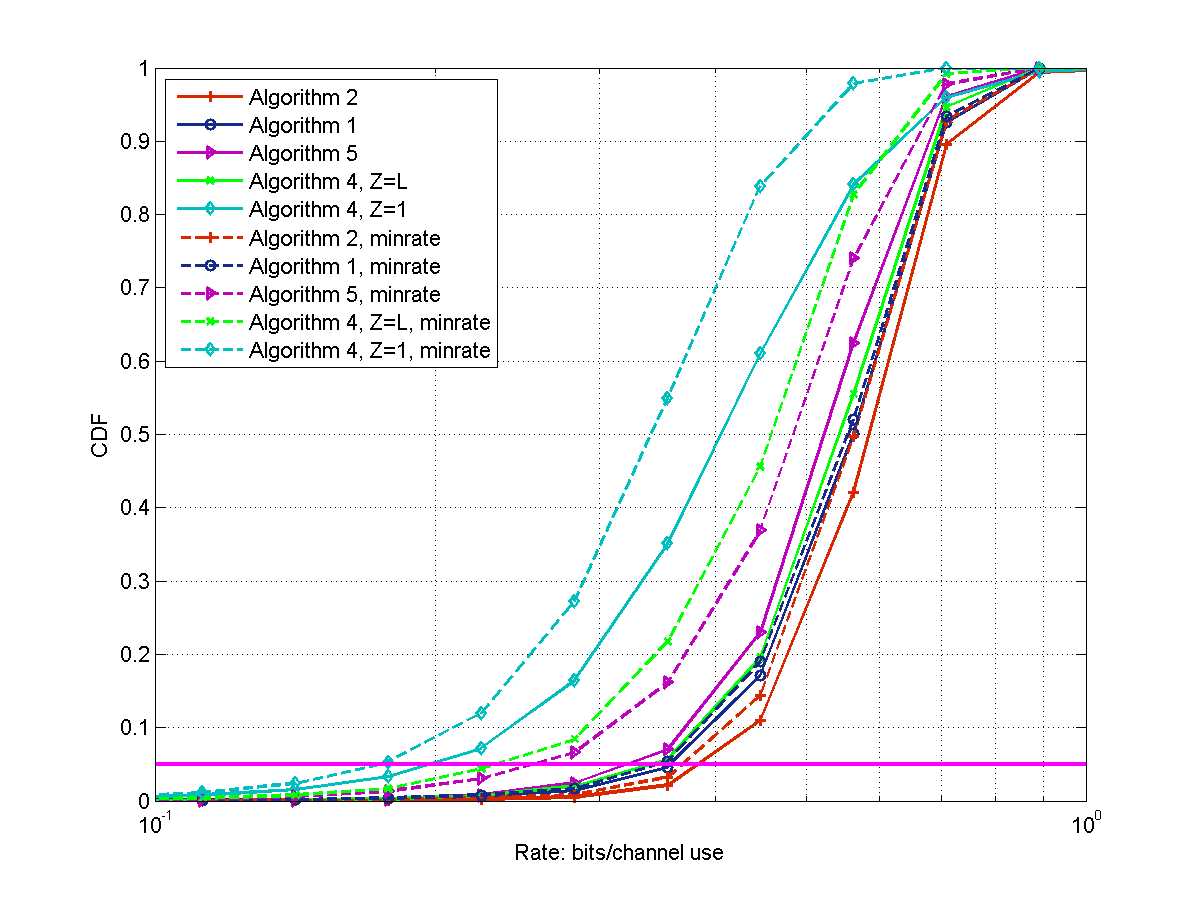}
\caption{The CDF of the achievable rates and minimum rates among all users for the proposed algorithms under the per-BS power constraints in a cellular network with $L=7$, $K=10$, and $M=64$. The solid curves refer to the CDF of rates of all users; the dashed curves refer to that of minimum rate among all users.}\label{fig-perBS2}
\end{figure}
\begin{table}
\centering
\caption{The relative performance on $R_{\mr{out}}$ of suboptimal algorithms with respect to the optimal algorithm for networks with $K=10, L=7$ under the per-BS power constraints.}\label{table-relate}
\begin{tabular}{|c|c|c|c|c|}
\hline
The number of BS antennas& Algorithm 5& Algorithm 4 ($Z=L$) & Algorithm 4 ($Z=1$) \\
\hline
$M=64 $ & $\frac{0.31}{0.36}=86.1\%$& $\frac{0.32}{0.36}=88.9\%$& $\frac{0.2}{0.36}=55.6\%$ \\
\hline
\end{tabular}
\end{table}
In Fig.~\ref{fig-perBS2}, we compare the algorithms under individual BS power constraints. All solid curves in Fig. \ref{fig-perBS2} denote the CDF of achievable rates of all uses. At the $5\%$ outage rate, Algorithm 2 achieves the rate
 $R_{\mr{out}}=0.38$ bits/channel use, while Algorithm 1 achieves $R_{\mr{out}}=0.36$ bits/channel use. Note that Algorithm 1 and 2 have individual BS power constraint and sum-power constraint, respectively. Performance of Algorithm 1 naturally upper-bounds that of Algorithm 2. However, we observe only $0.02$ bits/channel use performance degradation of applying individual BS power constraint. In Fig.~\ref{fig-perBS2}, we also show the CDF of the minimum rate among all users to verify the performance of proposed algorithms. All dashed curves denote corresponding performance. We can observe that, among the suboptimal Algorithm 4 with $Z=1$ and $Z=L$ as well as Algorithm 5, Algorithm 5 (purple dashed curve) achieve the best performance because of its power search procedure. However, when we turn to the CDF of achievable rates (solid curves), at the $5\%$ outage rate, Algorithm 4 with $Z=1$, $Z=L$, and Algorithm 5 achieves $R_{\mr{out}}=0.2$, $R_{\mr{out}}=0.32$, and $R_{\mr{out}}=0.31$, respectively. Algorithm 4 with $Z=L$ outperforms Algorithm 5. This is because our algorithms are targeted at maximizing the minimum rate of all users. For Algorithm 5, it sacrifices the rates of $95\%$ users to improve the performance of the worst $5\%$ users. The relative performance on $5\%$ outage rate of Algorithms 4 with $Z=L$ and $Z=1$ and Algorithm 5, compared to Algorithm 1, are summarized in Table.~\ref{table-relate}. As we can see, the achievable percentage is higher than the analyzed lowerbound $\frac{1}{L}$ in Subsection \ref{subsec-centalg}.

\begin{figure}[htb]
\includegraphics[scale=0.8]{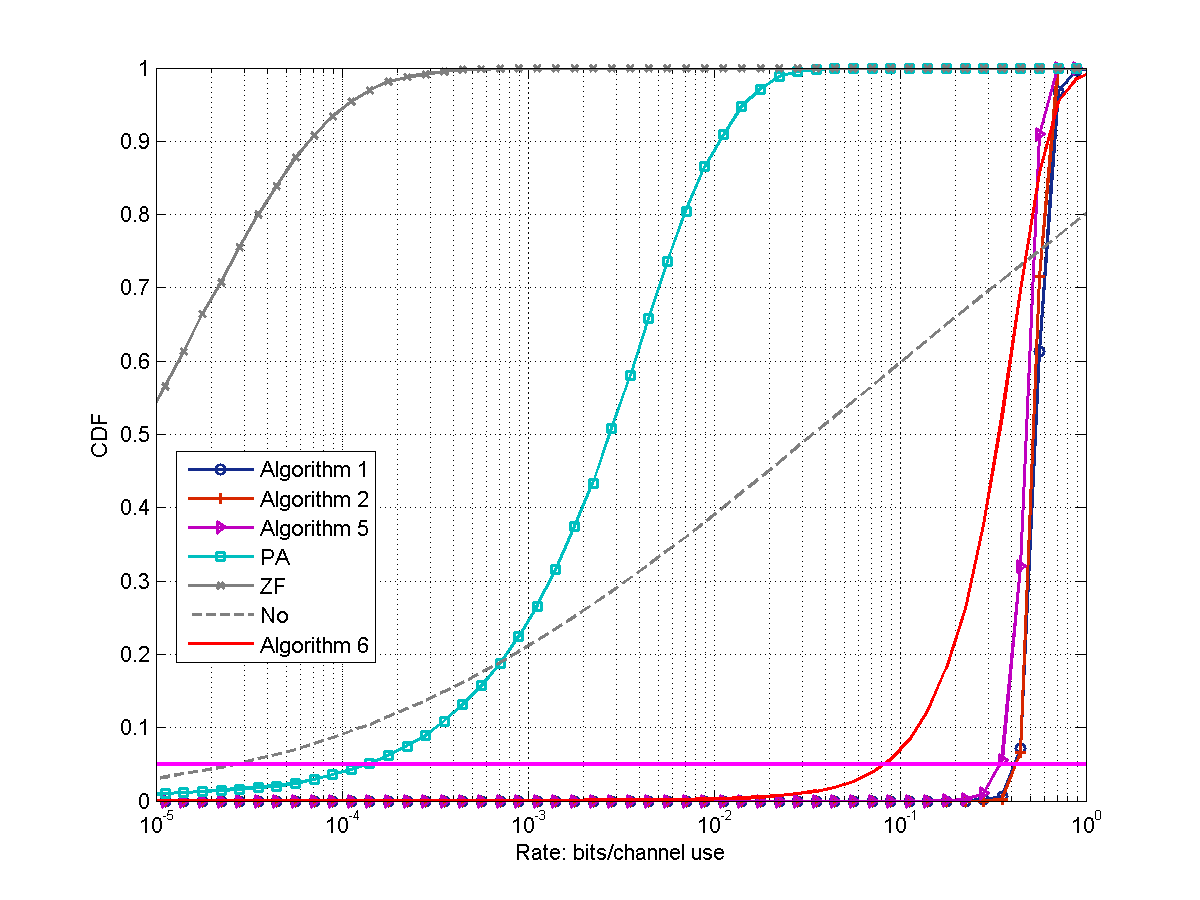}
\caption{The CDF of the achievable rates among all users for the related algorithms in a cellular network with $L=19$, $K=10$, and $M=64$. }\label{fig-case2}
\end{figure}

In Fig. \ref{fig-case2}, we present results for two-layer hexagonal cellular networks with $L=19$ cells. We first note that the gap between Algorithms 1 and 2 is reduced in $L=19$ compared to the case with $L=7$. As we allow more cells to cooperate, BSs are more likely to use equal power for the formulation with one sum-power constraint of all BSs. Further, compared to the case with one layer of cells, the achievable $R_{\mr{out}}$ of Algorithm 1 is increased from $0.35$ bits/channel use to $0.41$ bits/channel use. Similarly, we can also observe slight performance improvement of Algorithms 2 and 5. As more BSs are available for cooperation, the gain offered by multi-cell precoding has offset the additional interference incurred by a larger number of cells. Consequently, we observe performance improvement on $R_{\mr{out}}$ for Algorithms 1, 2, and 5 when increasing $L$ from seven cells to nineteen cells. However, for the PA algorithm, we observe performance degradation from $R_{\mr{out}}=4\times 10^{-4}$ in $L=7$ to $R_{\mr{out}}=1.1\times 10^{-4}$ in $L=19$. This implies the proposed algorithms are more effective in mitigating interference than the PA algorithm. 

Fig.~\ref{fig-case2} also includes the performance of Algorithm 6, which is a decentralized algorithm with per-BS power constraint as presented in Subsection \ref{subsec-decenalg}. In the simulation, Algorithm 6 uses Algorithm 5 to compute the precoding coefficients for the central BS. Note that all other algorithms in Fig. \ref{fig-case2} are centralized algorithms. Algorithm 6 achieves $0.084$ bits/channel use, which is around $\frac{0.084}{0.41}=20.5\%$ of Algorithm 1. The performance has significant improvement compared to the centralized PA algorithm, that achieves only $\frac{1.1\times 10^{-4}}{0.41}=0.027\%$ of Algorithm 1, because of local cooperation. The $5\%$ outage rate of Algorithm 6 is $\frac{0.084}{1.1\times 10^{-4}}=763.6$ times of that of the PA algorithm. However, we do not claim the optimality of Algorithm 6. Since the centralized Algorithm 5 can achieve $\frac{0.32}{0.41}=78.1\%$ performance of Algorithm 1, there is still a performance gap between the decentralized algorithm and the centralized Algorithms. The centralized algorithm needs to share data and large-scale fading coefficients globally. As the network grows large, the complexity of the centralized algorithms is prohibitive. Thus, further studies on decentralized LSFP algorithm is an important research topic.

\section{Conclusion}\label{sec-conclusion}
This paper has studied downlink multi-cell precoding algorithms using the large-scale fading coefficients for massive MIMO systems. We have proposed centralized precoding algorithms that require one hub connected to all cooperating cells. User symbols, large-scale fading coefficients, and precoding coefficients are exchanged through the central hub. The proposed multi-cell precoding algorithms can mitigate interference resulted from open access on the same channel when each BS is equipped with a practical large number of antennas.

We particularly considered precoding designs to maximize the minimum support rate among all users with individual BS power constraints. The formulated optimization problem is proved to be quasi-convex, thus optimally solvable using the bisection method together with feasibility checking. We also proposed suboptimal algorithms with reduced computation complexity by relaxing the Per-BS power constraint to sum-power constraint across all BSs, which is shown to be analytically tractable by uplink and downlink duality. We show through simulations that the proposed algorithms have an improvement on the $5\%$ outage rate for more than $1000$ times compared to the existing PA only algorithm, the ZF-LSFP, and the algorithm without LSFP.

\renewcommand{\baselinestretch}{1}
\bibliographystyle{ieeetran}
\bibliography{IEEEabrv,LSAS}

\end{document}